\documentclass[a4paper, 11pt]{article} 
\pdfoutput=1
\usepackage{jheppub}
\usepackage{xargs}
\usepackage[utf8]{inputenc}
\usepackage{graphicx}
\usepackage{csquotes}
\usepackage{xcolor}
\usepackage{amsmath}
\usepackage{amssymb}
\usepackage{mathtools}
\usepackage{dsfont}
\usepackage[low-sup]{subdepth}
\usepackage{siunitx}
\usepackage{enumitem}
\usepackage{hyperref}
\usepackage{cleveref}
\crefname{subsection}{subsection}{subsections}
\Crefname{subsection}{Subsection}{Subsections}
\usepackage{suffix} 
\usepackage{booktabs}
\usepackage{subcaption}
\usepackage{tabu}

\newcommand{\difspacing}[1]{\!\!{#1}\,\,} 
\newcommand{\pullint}{\!\!} 
\renewcommand{\d}[2][{}]{\mathrm{d}^{#1}{#2}}
\WithSuffix\newcommand{\d}*[2][{}]{\difspacing{\d[#1]{#2}}}
\newcommand{\dd}[3][{}]{\frac{\mathrm{d}^{#1}{#2}}{\mathrm{d}{#3}^{#1}}}
\newcommand{\pdd}[3][{}]{\frac{\partial^{#1}{#2}}{\partial{#3}^{#1}}}
\newcommandx{\fracd}[4][1={}, 4={}]{\frac{\mathrm{d}^{#1}{#2}}{{#3}^{#1}{#4}}}
\WithSuffix\newcommandx{\fracd}*[4][1={}, 4={}]{\difspacing{\fracd[#1]{#2}{#3}[#4]}}
\newcommand{\e}{\mathrm{e}}
\renewcommand{\i}{\mathrm{i}}
\newcommand{\ev}[1]{\langle{#1}\rangle}

\newcommand{\set}[1]{\{{#1}\}}
\renewcommand{\vec}[1]{\boldsymbol{#1}}
\newcommand{\sfrac}[2]{{#1}/{#2}} 

\usepackage{bm}
\newcommand{\vierer}[1]{\mathsf{#1}}
\newcommand{\vierervec}[1]{\bm{\vierer{#1}}}

\newcommand{\bra}[1]{\langle{#1}\vert}
\newcommand{\ket}[1]{\vert{#1}\rangle}

\newcommand{\bramidket}[3]{\langle{#1}\vert{#2}\vert{#3}\rangle}
\newcommand{\Bramidket}[3]{\left\langle{#1}\middle\vert{#2}\middle\vert{#3}\right\rangle} 

\newcommand{\Loperator}{L} 

\makeatletter
\DeclareRobustCommand*\uell{\mathpalette\@uell\relax}
\newcommand*\@uell[2]{
	\setbox0=\hbox{$#1\ell$}
	\setbox1=\hbox{\rotatebox{12}{$#1\ell$}}
	\dimen0=\wd0 \advance\dimen0 by -\wd1 \divide\dimen0 by 2
	\mathord{\lower 0.1ex \hbox{\kern\dimen0\unhbox1\kern\dimen0}}
}

\newcommand{\ispace}{\mathrm{s}} 
\newcommand{\Trans}{\mathrm{T}}
\newcommand{\Longi}{\mathrm{R}} 
\newcommand{\iTime}{\mathrm{U}}
\newcommand{\Mix}{\mathrm{M}}
\newcommand{\scal}{\mathrm{s}}
\newcommand{\longi}{\uell}
\newcommand{\enth}{\mathrm{w}}
\newcommand{\mix}{\mathrm{m}}
\newcommand{\sep}{\mathrm{r}}
\newcommand{\trans}{\mathrm{t}}
\newcommand{\itime}{\mathrm{u}} 

\newcommand{\all}{\mathrm{all}}

\begin{document}

\title{Tensor Decomposition for \linebreak Energy-Momentum Correlation Functions}

\author{Guy D. Moore and}
\author{Jonas Winter}
\affiliation{Institut f\"ur Kernphysik, Technische Universit\"at Darmstadt\\
Schlossgartenstra{\ss}e 2, D-64289 Darmstadt, Germany}
\emailAdd{jonas.winter@physik.tu-darmstadt.de}
\emailAdd{guy\_david.moore@tu-darmstadt.de}

\date{May 2026}

\abstract{
	We establish the general functional form of the energy-momentum-tensor two-point function in Euclidean coordinate space at zero and finite temperature.
	The full correlation function is first decomposed into its fundamental tensorial structures based on the remaining rotational symmetry.
	We use energy-momentum conservation to derive differential relations between the resulting component functions.
	Using these constraints, the full set of component functions of the correlator can finally be represented in the form of a smaller set of spectral functions.
	Finally, we show how to use these techniques for more efficient future lattice investigations.
}
\arxivnumber{arXiv:2606.10526}

\toccontinuoustrue 

\setcounter{tocdepth}{3} 

\maketitle

\section{Introduction}

Shear and bulk viscosity play key roles in our understanding of the quark-gluon plasma.
Experimentally, heavy-ion collisions are well described by hydrodynamics with a small but non-vanishing viscosity~\cite{
	PHENIX:2004vcz,
	BRAHMS:2004adc,
	PHOBOS:2004zne,
	STAR:2005gfr,
	Romatschke:2007mq,
	Luzum:2008cw,
	Bozek:2009dw,
	Schenke:2010nt,
	Song:2010aq,
	Schenke:2010rr,
	Schenke:2011bn,
	Heinz:2013th,
	Romatschke:2015gxa,
	Ryu:2017qzn,
	Bernhard:2019bmu,
	Shen:2020gef,
	JETSCAPE:2020mzn,
	JETSCAPE:2020shq,
	Gotz:2025wnv,
	Jaiswal:2025deb,
	Du:2025dpu%
}.
In comparison, \textsl{perturbative} treatments of hot quantum chromodynamics (QCD) predict a shear viscosity larger than that inferred from experiment, but with severe series-convergence problems
\cite{Arnold:2003zc,Ghiglieri:2018dib}.
This is not surprising, since the temperatures achieved in heavy-ion collisions are low enough that we should expect the hot strong-interaction matter to be strongly coupled.
Therefore non-perturbative tools, such as lattice calculations, are needed.
Karsch and Wyld showed how to determine the shear viscosity on the lattice by the analytical continuation of Euclidean energy-momentum tensor (EMT) two-point functions \cite{Karsch:1986cq}.
Since this seminal work, many authors have further explored and refined this approach~\cite{
	Nakamura:1996na,
	Sakai:1997ia,
	Aarts:2002vx,
	Nakamura:2004sy,
	Meyer:2007ic,
	Meyer:2007dy,
	Huebner:2008as,
	Meyer:2009jp,
	Meyer:2011gj,
	Mages:2015rea,
	Astrakhantsev:2017nrs,
	Kitazawa:2017qab,
	Astrakhantsev:2018oue,
	Borsanyi:2018srz,
	Altenkort:2022yhb%
}.
The general relation between hydrodynamic transport coefficients and stress-tensor correlation functions has been developed
\cite{Bhattacharyya:2007vjd,Baier:2007ix,Moore:2010bu,Kovtun:2018dvd}, meaning that lattice evaluations of EMT correlation functions can be used to extract several hydrodynamic transport coefficients.
Generally, these evaluations involve averaging the EMT correlation function over spatial separations at each fixed available Euclidean time separation.
The problem with this approach is that the correlation function decays rapidly with separation, but the fluctuations do not.
Therefore, integrating over separations includes a large volume which purely contributes noise without any signal.
Recently, the authors of Ref.~\cite{Altenkort:2021jbk,Altenkort:2022yhb} showed that one can achieve much more precise results by determining the full space-dependent correlation function, fitting its long-distance tail, and replacing the correlator at large distances with this tail fit.
This avoids including separations with poor signal to noise.

The cost of this approach is the need to determine the complete space-separation and time-separation dependence of the tensor's correlation function.
This correlation function is a rank-4 tensor, so its structure is somewhat involved, and it must be properly understood in order to carry out such a tail fitting.
For instance, the shear viscosity is often written in terms of the space $\vec r$ and Euclidean-time $\tau$ dependent EMT-EMT correlation function
$\int\d[3]{r} \langle T_{xy}(\tau,\vec{r})T_{xy}(0,\vec{0})\rangle$.
But this correlation function shows strong direction dependence: for small $\tau/r$, it is positive along the $z$-axis, negative along the $x$ and $y$ axes,
and positive again along the line $(x=y,z=0)$.
In order to fit the large-distance tail, it is therefore important to understand this complex angle dependence.
As we will show, it arises from the fact that this correlator actually represents an angle-dependent linear combination of three underlying correlation functions.
A better approach is to use all available data to extract these three functions separately, and to relate them to the shear viscosity.
The purpose of this paper is to explore the functional form of the EMT correlation function from general principles, in order to make its lattice extraction maximally efficient.

We begin by defining the two-point correlation function of the EMT $T_{\mu\nu}$:
\begin{align}
	G_{\mu\nu\rho\sigma}(\tau,\vec{r})&\coloneqq\ev{T_{\mu\nu}(\tau,\vec{r})T_{\rho\sigma}(0,\vec{0})}
	=\int\difspacing{\mathcal{D}\phi} \e^{-S[\phi]} T_{\mu\nu}(\tau,\vec{r})T_{\rho\sigma}(0,\vec{0}).
\end{align}
Here $\phi$ refers generically to the degrees of freedom of a Euclidean field theory.
As the EMT is a symmetric rank-2 tensor under rotations, its correlator is rank-4 under rotations and symmetric on the first two indices, the second two indices, and the exchange of the first pair with the second pair.
However, some rotations also change the direction of $(\tau,\vec{r})$; only rotations which leave the separation axis invariant (the little group) constrain the form of the correlator.
If one considers the theory in vacuum in $D$ spacetime dimensions, this leaves an $\mathrm{SO}(D-1)$ subgroup of the full $\mathrm{SO}(D)$ rotational symmetry, but at finite temperature, the finite and periodic temporal dimension further reduces the available symmetries to $\text{SO}(D-2)$.

Since the stress tensor is defined in terms of the variation of the action with respect to the spacetime metric, the above can also be written (provided $(\tau,\vec{r}) \neq (0,\vec{0})$, which we will assume throughout) as
\begin{align}
	G_{\mu\nu\rho\sigma}(\tau,\vec{r}) & =
	\frac{4\delta^2}{\delta g_{\mu\nu}(\tau,\vec r) \delta g_{\rho\sigma}(0,\vec{0})} \int\difspacing{\mathcal{D}\phi} \e^{-S[\phi,g]}.
\end{align}
As such, it is even under $\text{C}$-symmetry.
It is odd under reflections along the $(\tau,\vec r)$ axis if an odd number of $(\mu,\nu)$ and of $(\rho,\sigma)$ lie along that axis, and even under reflections across any axis orthogonal to the $(\tau,\vec r)$ direction.

As a product of two objects with ten independent components, the correlator is described by 100 components.
Symmetry under exchanging the first and second index pairs reduces this to 55 components.
However, the remaining rotational symmetry interrelates some of these components and tells some others to vanish.

For an optimal analysis of the correlator, especially when evaluating numerical data from e.g.\ lattice studies, these relations have to be taken into account without losing any physical information.
In this paper, we aim to decompose the full correlation function of two EMTs into its fundamental tensorial structures based on the relevant symmetries.
This tensor decomposition is performed in three scenarios: first in the vacuum, then at finite temperature when one averages of the temporal separation in the correlator and last at finite temperature with a general temporal separation.

The decomposition of the correlation function and the implications of energy-mo\-men\-tum conservation (EMC) in momentum space are well known: The \textsl{Minkowski, momentum-space} correlation function in vacuum\footnote{We thank Sangyong Jeon for useful conversations and Larry Yaffe for sharing unpublished notes.}
can be written in terms of five linearly independent tensors, which themselves are constructed out of the metric and the momentum vector.
EMC implies a polynomial constraint for the correlation function, which directly reduces the number of independent components to two.
At finite temperature there are 10 available tensor structures, and EMC provides a constraint which reduces this to 5.
In contrast, in coordinate space, EMC will enter as a differential constraint, and working out its consequences will be much more involved.

The next section will carry out the tensor decomposition of the EMT correlation function in terms of a minimum basis of orthonormal projectors.
Then,~\cref{sec:EMC} will show how these functions are interrelated by EMC.
The correlation function has a spectral decomposition which is useful for understanding the implications of these constraints and for understanding the large-distance limiting behavior of the correlation function.
We explore this in~\cref{sec:SD}.
In each of these sections, we first explore the simplest case -- vacuum field theory -- and then the more involved cases of finite temperature but averaged over the temporal direction, followed by finite temperature with both spatial and (Euclidean) temporal separations treated as non-vanishing.
It is therefore possible to read~\cref{subsec:EMC_zT,subsec:SD_zT} before reading~\cref{subsec:TD_fT_aTau}.

Finally, \cref{sec:app} briefly describes how our results can be used in lattice studies of the shear and bulk viscosity and of the \enquote{thermodynamic} transport coefficients discussed in Ref.~\cite{Kovtun:2018dvd}.

\section{Tensor Decomposition}\label{sec:TD}

\subsection{Zero Temperature}\label{subsec:TD_zT} 
In this section, the decomposition of the EMT correlator in $D$ Euclidean dimensions will be analyzed.
We treat $D$ as general so that we can easily specialize our results to both $D=4$ and to $D=3$ which are our main interest.
In this scenario, the temporal dimension is infinite and hence indistinguishable from the spatial dimensions. Hence, we denote the operator separation with a single symbol $\vierervec{r}\coloneqq(\tau,\vec{r})$.

\subsubsection{Decomposition of the EMT}\label{subsubsec:TD_zT_EMT_decomposition_4D}
In $D=4$ dimensions, we can take the separation $\vierervec{r}$ to lie in the temporal direction such that the EMT decomposes into the following structures:
\begin{itemize}[noitemsep]
	\item A traceless spatial tensor with 5 independent components: $T_{xy}$, $T_{xx}-T_{yy}$, etc.
	\item A vector consisting of 3 independent, one-space-one-time components: $T_{0x}$, etc.
	\item A traceless spatial scalar: $T_{00}-\frac{1}{3}(T_{xx}+T_{yy}+T_{zz})$
	\item The pure trace: $T_{\mu\mu}$
\end{itemize}
These four structures are a rank-2 tensor, a vector and two scalars
under $\text{SO}(D-1)$ rotations which preserve the $\vierervec{r}$ axis.
Because a tensor, a vector and a scalar transform differently under rotations, one can show that the correlations between these vanish.
The only correlation functions, which are allowed by rotational symmetry, are those within the same representation of the rotational group, i.e.\ one tensor-tensor, one vector-vector and three scalar-scalar (trace-trace, traceless-traceless and trace-traceless) correlation functions.
In total, there are five independent correlation functions.
Furthermore, considering how each component of $T_{\mu\nu}$ transforms under time reflection, the requirement that the theory be reflection-positive
\cite{Osterwalder:1973dx,Osterwalder:1974tc}
ensures that the correlator of an object which flips sign under time-reflection (here, the vector-vector correlator) is strictly negative, while the tensor-tensor correlator is positive and the $2\times 2$ matrix of scalar-scalar correlators has strictly non-negative eigenvalues.

This decomposition of the EMT is the same in $D$ dimensions, such that there are also only 5 correlation functions.
Furthermore, due to $\text{SO}(D)$ symmetry, this argument is still valid for any direction of the separation.

The goal this section is now to find the relevant tensorial structures which incorporate these rotational symmetries and to decompose the full correlator into these.

\subsubsection{Decomposition of the Correlation Function}\label{subsubsec:TD_zT_correlator_decomposition}
For the scalar part, there are two fundamental structures to be considered:
\begin{align}\label{eq:TD_zT_L_operator_definition}
	&\delta_{\mu\nu} &
	&\Loperator_{\mu\nu}\coloneqq D\hat{\vierer{r}}_{\mu}\hat{\vierer{r}}_{\nu} - \delta_{\mu\nu}
\end{align}
with $\hat{\vierer{r}}_{\mu}\coloneqq\frac{\vierer{r}_{\mu}}{\vierer{r}}$ and $\vierer{r}\coloneqq\vert\vierervec{r}\vert$.
With these two structures, one can define two projectors%
\footnote{
	\label{footnote:scalar_basis_choice}
	The choice here is not unique.
	We could choose $C^{\sep}_{\mu\nu}=\hat{\vierer{r}}_{\mu} \hat{\vierer{r}}_{\nu}$ and $C^{\trans}_{\mu\nu} = (\delta_{\mu\nu} - \hat{\vierer{r}}_{\mu} \hat{\vierer{r}}_{\nu})/\sqrt{D-1}$ instead.
	We choose the basis~\eqref{eq:TD_zT_scalar_rank_2_projectors}, here and in the following subsections, because it is more practical for our applications, even though it makes the consequences of energy-momentum conservation appear more complicated.
}
\begin{align}\label{eq:TD_zT_scalar_rank_2_projectors}
	C^{\scal}_{\mu\nu}&\coloneqq\frac{1}{\sqrt{D}}\delta_{\mu\nu} &
	C^{\longi}_{\mu\nu}&\coloneqq\frac{1}{\sqrt{D(D-1)}}\Loperator_{\mu\nu}
\end{align}
which project the EMT onto the two relevant scalars
found in the previous discussion, namely the trace ($\scal$) and the traceless, longitudinal part ($\longi$),
with a normalization convention chosen such that $C^{X}_{\mu\nu} C^{Y}_{\mu\nu} = \delta_{XY}$.
With these two two-index projectors, one can define three projectors
\begin{align}
	P^{\scal\scal}_{\mu\nu\rho\sigma}
	&\coloneqq C^{\scal}_{\mu\nu}C^{\scal}_{\rho\sigma}
	=\frac{1}{D}\delta_{\mu\nu}\delta_{\rho\sigma}
	\\
	P^{\longi\longi}_{\mu\nu\rho\sigma}
	&\coloneqq C^{\longi}_{\mu\nu}C^{\longi}_{\rho\sigma}
	=\frac{1}{D(D-1)}\Loperator_{\mu\nu}\Loperator_{\rho\sigma}
	\\
	P^{\scal\longi}_{\mu\nu\rho\sigma}
	&\coloneqq C^{\scal}_{\mu\nu}C^{\longi}_{\rho\sigma}+C^{\longi}_{\mu\nu}C^{\scal}_{\rho\sigma}
	=\frac{1}{D\sqrt{D-1}}\left[
		\Loperator_{\mu\nu}\delta_{\rho\sigma}
		+\delta_{\mu\nu}\Loperator_{\rho\sigma}
	\right]
\end{align}
which describe the decomposition of the correlator in the scalar sector.
Defining the product of two projectors,
\begin{align}
	(P^{XY} P^{ZW})_{\mu\nu\rho\sigma}
	&\coloneqq P^{XY}_{\mu\nu\alpha\beta} P^{ZW}_{\alpha\beta\rho\sigma}
\end{align}
These three projectors fulfill
\begin{subequations}
	\begin{align}
		P^{XX}P^{XX}&=P^{XX} \\
		P^{XX}P^{YY}&=0 \\
		P^{XY}P^{XY}&=P^{XX}+P^{YY} \\
		P^{XY}P^{XX}+P^{XX}P^{XY}&=P^{XY}
	\end{align}
\end{subequations}
for distinct $X,Y\in\set{\scal,\longi}$.
The two non-mixing projectors $P^{\scal\scal}$ and $P^{\longi\longi}$ are orthonormal and the mixing projector $P^{\scal\longi}$ is \emph{not} orthogonal to the other two.
The two projectors $P^{\scal\scal}$ and $P^{\longi\longi}$ and the projector $P^{\scal\longi}$ can be thought of as the on-diagonal and off-diagonal elements, respectively, of the $2\times2$ matrix of scalar projectors, which corresponds to the $2\times2$ matrix of scalar correlation functions.

Moving forward into the transverse sector, the structure
\begin{align}\label{eq:TD_zT_delta_transverse_definition}
	\delta^{\Trans}_{\mu\nu}&\coloneqq\delta_{\mu\nu}-\hat{\vierer{r}}_{\mu}\hat{\vierer{r}}_{\nu}
\end{align}
allows the construction of two projectors%
\footnote{Capital letters $\Longi,\Trans$ indicate that indices from the first and second pair are combined, while lower-case $\scal,\longi$ indicate that indices from a single pair are combined. $\Trans,\trans$ for transverse, $\Longi,\sep$ for radial, $\scal$ for scalar, $\longi$ for longitudinal, and (later) $\iTime,\itime$ for temporal and $\Mix,\mix$ for mixed radial-temporal.}
\begin{align}
	P^{\Longi\Trans}_{\mu\nu\rho\sigma}&\coloneqq\frac{1}{2}\left[
		\hat{\vierer{r}}_{\mu}\hat{\vierer{r}}_{\rho}\delta^{\Trans}_{\nu\sigma}
		+\hat{\vierer{r}}_{\nu}\hat{\vierer{r}}_{\rho}\delta^{\Trans}_{\mu\sigma}
		+\hat{\vierer{r}}_{\mu}\hat{\vierer{r}}_{\sigma}\delta^{\Trans}_{\nu\rho}
		+\hat{\vierer{r}}_{\nu}\hat{\vierer{r}}_{\sigma}\delta^{\Trans}_{\mu\rho}
	\right] \\
	P^{\Trans\Trans}_{\mu\nu\rho\sigma}
	&\coloneqq\frac{1}{2}(
		\delta^{\Trans}_{\mu\rho}\delta^{\Trans}_{\nu\sigma}
		+\delta^{\Trans}_{\mu\sigma}\delta^{\Trans}_{\nu\rho}
	)
	-\frac{1}{D-1}\delta^{\Trans}_{\mu\nu}\delta^{\Trans}_{\rho\sigma}
\end{align}
which correspond to the vector and rank-2 tensor part of the correlation function as they transform alike under rotation around the separation axis.
Both are orthogonal to each other and to the previously found projectors, and both are normalized properly.

Furthermore, one can check that the four orthogonal projectors form a complete set of projectors as they add up to an identity projector\footnote{The projector, which leaves all other projectors intact while obeying the relevant symmetries in its index structure.}
\begin{align}\label{eq:TD_zT_Pall}
	P^{\scal\scal}_{\mu\nu\rho\sigma}
	+P^{\longi\longi}_{\mu\nu\rho\sigma}
	+P^{\Longi\Trans}_{\mu\nu\rho\sigma}
	+P^{\Trans\Trans}_{\mu\nu\rho\sigma}
	&=P^{\mathrm{all}}_{\mu\nu\rho\sigma}
	\coloneqq\frac{1}{2}\left(
		\delta_{\mu\rho}\delta_{\nu\sigma}
		+\delta_{\mu\sigma}\delta_{\nu\rho}
	\right).
\end{align}
The full correlation function can be written in terms of these five projectors
\begin{align}
	\begin{split}
		G_{\mu\nu\rho\sigma}(\vierervec{r})
		&
		\begin{aligned}[t]
			{}=G^{\Trans\Trans}(\vierer{r}^{2})P^{\Trans\Trans}_{\mu\nu\rho\sigma}(\hat{\vierervec{r}})
			&+G^{\Longi\Trans}(\vierer{r}^{2})P^{\Longi\Trans}_{\mu\nu\rho\sigma}(\hat{\vierervec{r}}) \\
			+G^{\scal\scal}(\vierer{r}^{2})P^{\scal\scal}_{\mu\nu\rho\sigma}(\hat{\vierervec{r}})
			&+G^{\longi\longi}(\vierer{r}^{2})P^{\longi\longi}_{\mu\nu\rho\sigma}(\hat{\vierervec{r}})
			+G^{\scal\longi}(\vierer{r}^{2})P^{\scal\longi}_{\mu\nu\rho\sigma}(\hat{\vierervec{r}})
		\end{aligned}
	\end{split}
\end{align}
with five component functions $G^{\Trans\Trans}$, $G^{\Longi\Trans}$, $G^{\scal\scal}$, $G^{\longi\longi}$ and $G^{\scal\longi}$ depending only on the separation\footnote{Or equivalently on $\vierer{r}^{2}$, which is more convenient in some situations.} $\vierer{r}$, because the angular dependence $\hat{\vierervec{r}}$ is absorbed into the projectors.
These five functions can be extracted by projecting the respective projector against the full correlator
\begin{align}
	G^{X}(\vierer{r}^{2})&=\frac{1}{M^{X}}P^{X}_{\mu\nu\rho\sigma}(\hat{\vierervec{r}})G_{\mu\nu\rho\sigma}(\vierervec{r})
\end{align}
for $X\in\mathcal{I}_{0}\coloneqq\set{\Trans\Trans,\Longi\Trans,\scal\scal,\longi\longi,\scal\longi}$,
where $M^{X}\coloneqq P^{X}_{\mu\nu\rho\sigma} P^{X}_{\mu\nu\rho\sigma}$ is the multiplicity of a projector, listed in~\cref{tab:TD_zT_multiplicites}.

\begin{table}[htb]
	\centering
	\begin{tabular}{@{} *{4}{c} @{}}
		\toprule
		$X$ & $M^{X}$ & $M^{X}(D=4)$ & $M^{X}(D=3)$ \\
		\midrule
		$\Trans\Trans$ & $\frac{(D{+}1)(D{-}2)}{2}$ & 5 & 2 \\ \addlinespace
		$\Longi\Trans$ & $D{-}1$ & 3 & 2 \\
		$\scal\scal$ & 1 & 1 & 1 \\
		$\longi\longi$ & 1 & 1 & 1 \\
		$\scal\longi$ & 2 & 2 & 2 \\
		\bottomrule
	\end{tabular}
	\caption{Projector multiplicities in $D$, $4$, and $3$ dimensions.}
	\label{tab:TD_zT_multiplicites}
\end{table}

The multiplicity $M^{\scal\longi}=2$ reflects the fact that there are two off-diagonal elements in the $2\times2$ matrix of scalar correlators.
In $D=4$, the multiplicities of the tensor mode $M^{\Trans\Trans}=5$ and of the vector mode $M^{\Longi\Trans}=3$ reflect the numbers of components of the known decomposition of the EMT, see~\cref{subsubsec:TD_zT_EMT_decomposition_4D}.
In $D=3$, the multiplicities of the tensor mode $M^{\Trans\Trans}=2$ and of the vector mode $M^{\Longi\Trans}=2$ are equivalent to the number of polarizations of a graviton and a photon, respectively.

So far, it appears that $G_{\mu\nu\rho\sigma}$ is expressed in terms of five independent component functions.
However, EMC will reduce this to two true independent functions.
In coordinate space, the inter-relations are differential, rather than algebraic, and have non-trivial implications.
So as not to break the logical flow of the presentation, we will return to this issue in~\cref{subsec:EMC_zT} after investigating the tensor decomposition for the finite-temperature case in the sections.

\subsection{Finite Temperature --- Averaging over time}\label{subsec:TD_fT_aTau}

At finite temperature, the Euclidean time direction $\tau$ is periodic with period $\beta = \frac 1T$.
This distinguishes the $\tau$ direction from the three spatial directions $\vec{r}$, so the correlation function is $G_{\mu\nu\rho\sigma}(\tau,\vec{r})$.
For e.g.\ the computation of second-order transport coefficients, see~\cite{Kovtun:2018dvd}, we actually want the average over $\tau$:
\begin{align}
	G_{\mu\nu\rho\sigma}(\vec{r})&=\int_{0}^{\beta}\d*{\tau}\ev{T_{\mu\nu}(\tau, \vec{r})T_{\rho\sigma}(0,\vec{0})}.
\end{align}
It behaves very much like the $D=3$ vacuum case \emph{except} that the EMT is a $4\times4$ symmetric matrix.

For the formulation of the projectors, the unit time vector $\vec{u}$ is introduced, which defines the equilibrium rest frame of the system via $u^{\mu}=\delta^{\mu0}$.
In addition to the previously appearing structures $\delta$ and $\delta^{\Trans}$, their purely-spatial versions are given by
\begin{align}
	\delta^{\ispace}_{\mu\nu}
	&\coloneqq\delta_{\mu\nu}-u_{\mu}u_{\nu}
	&
	\delta^{\ispace\Trans}_{\mu\nu}
	&\coloneqq\delta^{\ispace}_{\mu\nu}-\hat{r}_{\mu}\hat{r}_{\nu}
	=\delta_{\mu\nu}-u_{\mu}u_{\nu}-\hat{r}_{\mu}\hat{r}_{\nu}
\end{align}
and project to the purely-spatial trace and the purely-spatial transverse part.

In analogy to the previous section, the projectors%
\footnote{Note that the meaning of $\Trans$ is different than in the last subsection:  it now refers to indices which are transverse
both with respect to the spatial separation and with respect to time.}
\begin{align}
	P^{\Trans\Trans}_{\mu\nu\rho\sigma}&\coloneqq\frac{1}{2}\left(
		\delta^{\ispace\Trans}_{\mu\rho}\delta^{\ispace\Trans}_{\nu\sigma}
		+\delta^{\ispace\Trans}_{\mu\sigma}\delta^{\ispace\Trans}_{\nu\rho}
		-\delta^{\ispace\Trans}_{\mu\nu}\delta^{\ispace\Trans}_{\rho\sigma}
	\right) \label{eq:TD_fT_aTau_P_TT} \\
	P^{\Longi\Trans}_{\mu\nu\rho\sigma}&\coloneqq\frac{1}{2}\left[
		\hat{r}_{\mu}\hat{r}_{\rho}\delta^{\ispace\Trans}_{\nu\sigma}
		+\hat{r}_{\mu}\hat{r}_{\sigma}\delta^{\ispace\Trans}_{\nu\rho}
		+\hat{r}_{\nu}\hat{r}_{\rho}\delta^{\ispace\Trans}_{\mu\sigma}
		+\hat{r}_{\nu}\hat{r}_{\sigma}\delta^{\ispace\Trans}_{\mu\rho}
	\right] \label{eq:TD_fT_aTau_P_RT}
\end{align}
correspond to the purely-spatial rank-2 tensor and vector, respectively.
Both are orthogonal to each other and are properly normalized.
As they are structures in the three-dimensional subspace, they each have two independent components as seen in the previous section.

Furthermore, one can construct two projectors which have two temporal and two spatial indices, namely
\begin{align}
	P^{\iTime\Trans}_{\mu\nu\rho\sigma}&\coloneqq\frac{1}{2}\left[
		\delta^{\ispace\Trans}_{\mu\rho}u_{\nu}u_{\sigma}
		+\delta^{\ispace\Trans}_{\mu\sigma}u_{\nu}u_{\rho}
		+\delta^{\ispace\Trans}_{\nu\rho}u_{\mu}u_{\sigma}
		+\delta^{\ispace\Trans}_{\nu\sigma}u_{\mu}u_{\rho}
	\right] \label{eq:TD_fT_aTau_P_UT}\\
	P^{\iTime\Longi}_{\mu\nu\rho\sigma} &\coloneqq
	\frac{1}{2} \left[
		\hat r_{\mu} \hat r_{\rho}   u_{\nu} u_{\sigma}
		+ \hat r_{\mu} \hat r_{\sigma} u_{\nu} u_{\rho}
		+ \hat r_{\nu} \hat r_{\rho}   u_{\mu} u_{\sigma}
	+ \hat r_{\nu} \hat r_{\sigma} u_{\mu} u_{\rho} \right]
	\label{eq:TD_fT_aTau_P_UL}
\end{align}
corresponding to a vector and a scalar, respectively.
Both are again orthogonal to each other and to the previous two projectors, and both are properly normalized.

The scalar sector is extended by one additional rank-2 projector $C^{\enth}$,
\begin{align}
	C^{\scal}_{\mu\nu}&\coloneqq\frac{1}{2}\delta_{\mu\nu} &
	C^{\longi}_{\mu\nu}&\coloneqq \frac{1}{\sqrt{6}} \left( 3\hat{r}_{\mu}\hat{r}_{\nu}- \delta^{\ispace}_{\mu\nu}\right) &
	C^{\enth}_{\mu\nu}&\coloneqq\frac{1}{\sqrt{12}}\left( 3u_{\mu}u_{\nu} -\delta^{\ispace}_{\mu\nu} \right),
\end{align}
which projects the EMT to the enthalpy%
\footnote{The enthalpy density in Minkowski signature is $w=\varepsilon+P=T_{00}+\frac{1}{3} T_{ii}$, but in Euclidean signature the sign of $T_{00}$ is reversed.}
density
$C^{\enth}_{\mu\nu}T_{\mu\nu}\propto T_{ii}-3T_{00}=3w$.
The three form an orthonormal set of projectors and can be used to construct six scalar projectors, three diagonal and three off-diagonal, via
\begin{align}
	\label{eq:TD_fT_aTau_scalar_projectors}
	P^{XX}_{\mu\nu\rho\sigma}&\coloneqq C^{X}_{\mu\nu}C^{X}_{\rho\sigma} &
	P^{XY}_{\mu\nu\rho\sigma}&\coloneqq C^{X}_{\mu\nu}C^{Y}_{\rho\sigma}+C^{Y}_{\mu\nu}C^{X}_{\rho\sigma}
\end{align}
for $X\neq Y\in\set{\scal,\longi,\enth}$.
They fulfil a set of relations, similar to the relations found in the previous section, especially orthonormalization of the diagonal projectors:
\begin{subequations}\label{eq:TD_fT_aTau_scalar_projectors_relations}
	\begin{align}
		P^{XX}P^{XX}&=P^{XX} \label{eq:finite_T_averaged_t_projectors_scalar_diag_normalization}\\
		P^{XX}P^{YY}&=0 \\
		P^{XX}P^{XY}+P^{XY}P^{XX}&=P^{XY} \\
		P^{XX}P^{YZ}+P^{YZ}P^{XX}&=0 \\
		P^{XY}P^{XY}&=P^{XX}+P^{YY} \\
		P^{XY}P^{YZ}+P^{YZ}P^{XY}&=P^{XZ}
	\end{align}
\end{subequations}
for distinct $X,Y,Z\in\set{\scal,\longi,\enth}$.

With the projectors $P^{\Trans\Trans}$, $P^{\Longi\Trans}$, $P^{\iTime\Trans}$, $P^{\iTime\Longi}$ and the other six scalar projectors, the full correlation function can be written as
\begin{align}
	\label{eq:TD_fT_aTau_G}
	G_{\mu\nu\rho\sigma}(\vec{r})&=\sum_{X\in\mathcal{I}_{\bar{\tau}}}
	G^{X}(r^{2})P^{X}_{\mu\nu\rho\sigma}(\hat{\vec{r}})
	\\
	G^{X}(r^{2})&=\frac{1}{M^{X}}P^{X}_{\mu\nu\rho\sigma}(\hat{\vec{r}})G_{\mu\nu\rho\sigma}(\vec{r})
	\label{eq:fT_GXfromPG}
\end{align}
with $X\in\mathcal{I}_{\bar{\tau}}\coloneqq\set{\Trans\Trans,\Longi\Trans,\iTime\Trans, \iTime\Longi,\scal\scal,\longi\longi,\enth\enth,\scal\longi,\scal\enth,\longi\enth}$.
The multiplicities $M^{X}$ are $2$ for the tensorial component $\Trans\Trans$, $2$ for the vector components $\Longi\Trans$ and $\iTime\Trans$, $1$ for the diagonal scalar components and $\iTime\Longi$, and $2$ for the off-diagonal scalar components, matching the results from the previous subsection in~\cref{tab:TD_zT_multiplicites} for $D=3$ as space is three-dimensional in this subsection.

Defining $T^{X}=C^{X}_{\mu\nu}T_{\mu\nu}$ for $X\in\set{\scal,\longi,\enth}$, the scalar sector $(\scal,\longi,\enth)$ consists of every correlation of one of $(T^{\scal},T^{\longi},T^{\enth})$ at $(\tau,\vec{r})$ with another at $(0,\vec{0})$.
Furthermore, reflection positivity ensures
\begin{itemize}[noitemsep]
	\item $G^{\Longi\Trans}<0$ and $G^{\iTime\Longi}<0$,
	\item $G^{\Trans\Trans}>0$ and $G^{\iTime\Trans}>0$,
	\item the 
		$3\times3$ matrix of correlation functions over $(\scal,\longi,\enth)$ has positive eigenvalues.
\end{itemize}

EMC will provide 5 differential relations between these 10 functions, but we will postpone the complete investigation of these relations to~\cref{subsec:EMC_fT_aTau} until we have considered the most general case of tensor decomposition.

\subsection{Finite Temperature --- General time}\label{subsec:TD_fT_gTau}

In this section, we discuss the decomposition of the correlation function of two EMTs in a finite temporal dimension
\begin{align}
	G_{\mu\nu\rho\sigma}(\tau,\vec{r})&=\ev{T_{\mu\nu}(\tau,\vec{r})T_{\rho\sigma}(0,\vec{0})},
\end{align}
when there is no integration over the temporal separation $\tau$.
Again, $\vec{r}$ denotes the spatial separation only.
Correlation functions of this form are relevant for e.g.\ the calculation of transport coefficients like the shear and bulk viscosities.

Similarly to the previous section, most generators of vacuum $\text{SO}(4)$ symmetry either transform $\vec{r}$ or $\vec{u}$ and therefore do not constrain the form of $G_{\mu\nu\rho\sigma}(\tau,\vec{r})$ at fixed $\tau,\vec{r}$.
Only an $\text{SO}(2)$ subgroup leaves both vectors invariant, and time-reflection symmetry now only relates correlator values at $+\tau$ and $-\tau$.
Hence, the only guiding principle for construction the tensor decomposition of the upper tensor is the symmetry under rotations around the $\vec{r}$ axis.

\subsubsection{Decomposition of the EMT}\label{subsubsec:TD_fT_gTau_EMT_decomposition}

To understand this issue better, we discuss the decomposition of a single EMT in this scenario and its transformation behavior under rotations around $\vec{r}\parallel\vec{e}_{z}$.
\begin{itemize}[noitemsep]
	\item The components $(T_{xy},T_{xx}-T_{yy})$ form a doublet of spin-$2$ components.
		They are directly related by rotations around the $z$ axis and correspond to a spatial, traceless tensor.
	\item The components $(T_{xz},T_{yz})$ and $(T_{0x},T_{0y})$ form two doublets of spin-$1$ components and correspond to two vectors with respect to the $z$-rotations.
	\item The components $T_{00}$, $T_{0z}$, $T_{zz}$ and $T_{xx}+T_{yy}$ are each spin-$0$ about the $z$-axis and correspond to every scalar under rotations around the $z$ axis.
		The first, third and fourth already appeared in the previous section related to scalars, however the second one was a component of one of the spatial vectors.
\end{itemize}
In total, the EMT decomposes into one tensor, two vectors and four scalars.

As discussed before, rotational symmetry arguments restrict correlations only within a representation of the little group.
Hence, one expects a single correlation function between the tensor components, three\footnote{Two non-mixing contributions and one mixing contribution, forming a symmetric $2\times 2$ matrix.} correlation functions between the two vector components and ten\footnote{Four non-mixing contributions and six mixing contributions, forming a symmetric $4\times 4$ matrix.} correlation functions between the four scalar components.
In total, there will be fourteen individual functions describing the full correlation function and again, this argumentation holds for every choice of coordinates (not only $\vec{e}_{z}\parallel\vec{r}$) due to $\text{SO}(3)$ spatial rotational symmetry.

\subsubsection{Decomposition of the Correlation Function}\label{subsubsec:TD_fT_gTau_correlator_decomposition}

The spatial, traceless tensor is the same as in the previous section and hence, it also corresponds to the same projector~\eqref{eq:TD_fT_aTau_P_TT}
\begin{align*} 
	P^{\Trans\Trans}_{\mu\nu\rho\sigma}&=\frac{1}{2}\left(
		\delta^{\ispace\Trans}_{\mu\rho}\delta^{\ispace\Trans}_{\nu\sigma}
		+\delta^{\ispace\Trans}_{\mu\sigma}\delta^{\ispace\Trans}_{\nu\rho}
		-\delta^{\ispace\Trans}_{\mu\nu}\delta^{\ispace\Trans}_{\rho\sigma}
	\right).
\end{align*}
Moreover, there are three structures which are half transversal, and half in the symmetry breaking directions ($\vec{u}$ or $\vec{r}$) or a mix of both.
These are
\begin{align}
	P^{\Longi\Trans}_{\mu\nu\rho\sigma}&=\frac{1}{2}\left[
		\hat{r}_{\mu}\hat{r}_{\rho}\delta^{\ispace\Trans}_{\nu\sigma}
		+\hat{r}_{\mu}\hat{r}_{\sigma}\delta^{\ispace\Trans}_{\nu\rho}
		+\hat{r}_{\nu}\hat{r}_{\rho}\delta^{\ispace\Trans}_{\mu\sigma}
		+\hat{r}_{\nu}\hat{r}_{\sigma}\delta^{\ispace\Trans}_{\mu\rho}
	\right]
	\\
	P^{\Mix\Trans}_{\mu\nu\rho\sigma}&=\phantom{\frac{1}{2}}\left[ 
		\hat{r}_{(\mu}u_{\rho)} \delta^{\ispace\Trans}_{\nu\sigma}
		+\hat{r}_{(\mu}u_{\sigma)} \delta^{\ispace\Trans}_{\nu\rho}
		+\hat{r}_{(\nu}u_{\rho)} \delta^{\ispace\Trans}_{\mu\sigma}
		+\hat{r}_{(\nu}u_{\sigma)} \delta^{\ispace\Trans}_{\mu\rho}
	\right]
	\\
	P^{\iTime\Trans}_{\mu\nu\rho\sigma}&=\frac{1}{2}\left[
		u_{\mu}u_{\rho}\delta^{\ispace\Trans}_{\nu\sigma}
		+u_{\mu}u_{\sigma}\delta^{\ispace\Trans}_{\nu\rho}
		+u_{\nu}u_{\rho}\delta^{\ispace\Trans}_{\mu\sigma}
		+u_{\nu}u_{\sigma}\delta^{\ispace\Trans}_{\mu\rho}
	\right].
\end{align}
Here we use the notation
\begin{align}
	V_{(\mu} U_{\nu)} &\coloneqq \frac{1}{2} \left( V_{\mu} U_{\nu} + V_{\nu} U_{\mu} \right)
\end{align}
to indicate the index-ordering-averaged combination of two vectors.
They are orthogonal to the previously defined tensorial projector $P^{\Trans\Trans}$ and fulfill the following relations
\begin{subequations}
	\begin{align}
		P^{X\Trans}P^{X\Trans}&=P^{X\Trans} &
		X&\in\set{\Longi,\iTime}
		\\
		P^{\Longi\Trans}P^{\iTime\Trans}&=0
		\\
		P^{X\Trans}P^{\Mix\Trans}+P^{\Mix\Trans}P^{X\Trans}&=P^{\Mix\Trans} &
		X&\in\set{\Longi,\iTime}
		\\
		P^{\Mix\Trans}P^{\Mix\Trans}&=P^{\Longi\Trans}+P^{\iTime\Trans}
	\end{align}
\end{subequations}
in analogy to the relations from the scalar sector, see~\cref{eq:TD_fT_aTau_scalar_projectors_relations}.

The scalar sector is extended relative to the previous subsection by one additional rank-2 projector:
\begin{subequations}
	\begin{align}
		C^{\scal}_{\mu\nu}&=\frac{1}{2}\delta_{\mu\nu} &
		C^{\longi}_{\mu\nu}&=\frac{1}{\sqrt{6}}\left( 3\hat{r}_{\mu}\hat{r}_{\nu}- \delta^{\ispace}_{\mu\nu}\right) \\
		C^{\enth}_{\mu\nu}&=\frac{1}{\sqrt{12}}\left(3u_{\mu}u_{\nu} - \delta^{\ispace}_{\mu\nu} \right) &
		C^{\mix}_{\mu\nu}&=\sqrt{2}\, \hat{r}_{(\mu}u_{\nu)}.
	\end{align}
\end{subequations}
Here, $\mix$ denotes the mixing of one index projecting on $\vec{r}$ and one index projecting on $\vec{u}$.
These build four non-mixing and six mixing projectors in strict analogy to~\cref{eq:TD_fT_aTau_scalar_projectors},
which multiply each other exactly as in~\cref{eq:TD_fT_aTau_scalar_projectors_relations}.

Note that the projector $P^{\mix\mix}$ already appeared in the last subsection as $P^{\iTime\Longi}$ because it is even under time-reflection.
But now the mixing projectors $P^{\scal\mix},P^{\longi\mix}$ and $P^{\enth\mix}$ represent cross-correlations between $\mix$ and the other scalars.
These three, together with $P^{\Mix\Trans}$, are four new, odd-in-$u_{\mu}$ projectors whose associated correlators are odd under $\tau \to -\tau$ and therefore vanish under $\tau$ averaging.
In addition, the component functions now depend separately on $\tau$ and $r^{2}$.

With the tensorial projector $P^{\Trans\Trans}$, the three vector projectors $P^{\Longi\Trans}$, $P^{\Mix\Trans}$ and $P^{\iTime\Trans}$ and the ten scalar projectors, the full correlation function can be written as
\begin{subequations}
	\begin{align}
		G_{\mu\nu\rho\sigma}(\tau,\vec{r})&=\sum_{X\in\mathcal{I}_{\tau}}G^{X}(\tau,r^{2})P^{X}_{\mu\nu\rho\sigma}(\hat{\vec{r}})
		\\
		G^{X}(\tau,r^{2})&=\frac{1}{M^{X}}P^{X}_{\mu\nu\rho\sigma}(\hat{\vec{r}})G_{\mu\nu\rho\sigma}(\tau,\vec{r})
		\label{eq:aTauGXfromPG}
	\end{align}
\end{subequations}
with $X\in\mathcal{I}_{\tau}\coloneqq\set{ \Trans\Trans, \Longi\Trans, \Mix\Trans, \iTime\Trans, \scal\scal, \longi\longi, \enth\enth, \mix\mix, \scal\longi, \scal\enth, \scal\mix, \longi\enth, \longi\mix, \enth\mix}$.
The multiplicities $M^{X}$ are again $2$ for the tensorial components $\Trans\Trans$, $2$ for the vector components $\Longi\Trans$ and $\iTime\Trans$, $4$ for vector component $\Mix\Trans$\footnote{The multiplicity $M^{\Mix\Trans}=2\cdot2=4$ stems from it being a vector (the first $2$) and from it being the off-diagonal component in the $2\times2$ matrix of correlations in the vector sector (the second $2$).}, $1$ for the diagonal scalar components and $2$ for the off-diagonal scalar components, again matching the previously found results.

As noted before, the correlation functions for which $X$ contains exactly one $\mix$ are odd under $\tau \to -\tau$.
They therefore vanish when we average over $\tau$.
In addition, they approach zero when $r \gg \tau$:
\begin{align}
	G^{\Mix\Trans},G^{\scal\mix},G^{\longi\mix},G^{\enth\mix}&\xrightarrow{\text{$\tau$-average or $\sfrac{r}{\tau}\gg1$}}0.
\end{align}
Causality actually ensures that the approach to zero at large $r$ is at least as fast as $\exp(-2\pi r T)$.
In either of these cases, the general case considered here becomes equivalent to the $\tau$-averaged case considered in~\cref{subsec:TD_fT_aTau}.

\section{Implications of Energy-Momentum Conservation}\label{sec:EMC}
In this section, we explore the implications of EMC for the tensor decomposition of the correlation functions of two EMTs.

As mentioned in the introduction, the interrelations between the different components emerging from EMC are well known in momentum space and are much simpler than their position space counterpart, because they are only algebraic and thereby remove some of the tensorial structures directly.

To see this, the vacuum, momentum-space correlation function $G_{\mu\nu\rho\sigma}(\vierervec{k})$ can be expressed in terms of five fundamental tensor structures in analogy to~\cref{subsec:TD_zT}, which themselves are constructed out of $\delta$ and $\vierervec{k}$ components (instead of $\vierervec{r}$ components).
In momentum space, the conservation of energy and momentum is expressed via $\vierer{k}^{\mu}T_{\mu\nu}=0$ and hence $\vierer{k}^{\mu}G_{\mu\nu\rho\sigma}=0$ holds true for the full correlation function.
This relation implies the direct cancellation of one of the five components while three components are equal up to factors.
Ultimately, the relations implied by EMC reduce the number of independent components directly to two and the full correlation function can be expressed as
\begin{align}\label{eq:EMC_momentum_space}
	G_{\mu\nu\rho\sigma}(\vierervec{k}) & \overset{T=0}{=}
	G_{\text{traceless}}(\vierer{k}^{2}) \left( g^{\Trans}_{\mu\sigma} g^{\Trans}_{\nu\rho} + g^{\Trans}_{\mu\rho} g^{\Trans}_{\nu\sigma} - \frac{1}{3} g^{\Trans}_{\mu\nu} g^{\Trans}_{\rho\sigma} \right)
	+ G_{\text{trace}}(\vierer{k}^{2}) \: g^{\Trans}_{\mu\nu}  g^{\Trans}_{\rho\sigma} ,
\end{align}
with $g^{\Trans}_{\mu\nu} \coloneqq g_{\mu\nu} - \hat{\vierer{k}}_{\mu} \hat{\vierer{k}}_{\nu}$ and $g_{\mu\nu}$ being the metric.
To our knowledge this decomposition was first comprehensively discussed in Ref.~\cite{Kovtun:2005ev}, which also showed how it generalizes at finite temperature.
See also~\cite{Brandt:1998hd,Jeon:2025kem}.

In position space, however, the relations implied by EMC will turn out to be differential, and they only interrelate some of the correlation functions instead of erasing them.
The starting point for this discussion is EMC for the correlator
\begin{align}\label{eq:EMT_correlator_conservation}
	0&=
	\partial^{\vierervec{r}}_{\mu}\ev{T_{\mu\nu}(\vierervec{r})T_{\rho\sigma}(\vierervec{0})}
	=\partial^{\vierervec{r}}_{\mu}G_{\mu\nu\rho\sigma}(\vierervec{r}).
\end{align}
Note that strictly speaking this relation only holds when $\vierervec{r} \neq 0$; at the point $\vierervec{r} = 0$ it is spoiled by contact terms.
Since our primary interest is the case $\vierervec{r} \neq 0$ and we will not consider spacetime-integrals over the resulting equations, the presence of such contact terms will not affect the following discussion.

As a structure with three free indices $\nu$, $\rho$, and $\sigma$, one can construct all three-index structures $F^{(i)}_{\nu\rho\sigma}$, which respect the remaining index symmetry and especially the remaining rotational symmetry regarding the space-time structure of the three scenarios from the previous three sections.
The r.h.s.\ of~\cref{eq:EMT_correlator_conservation} can be decomposed in terms of these three-index structures:
\begin{align}\label{eq:EMT_correlator_conservation_as_Fs}
	0&=\partial^{\vierervec{r}}_{\mu}G_{\mu\nu\rho\sigma}(\vierervec{r})
	=\sum_{i}c_{i}F^{(i)}_{\nu\rho\sigma}
\end{align}
with coefficients $c_{i}$ being linear combinations of the correlation functions $G^{X}$ and their derivatives.
For this expression to hold true for every separation $\vec{r}$, each of these coefficients must vanish, yielding a differential relation between the correlation functions for each tensorial structure $F^{(i)}_{\nu\rho\sigma}$.

\subsection{Zero Temperature}\label{subsec:EMC_zT}
In vacuum with $\mathrm{O}(D)$ symmetry, there are three tensorial structures
\begin{align}\label{eq:EMC_zT_F_tensors}
	F^{(1)}_{\nu\rho\sigma}(\hat{\vierervec{r}})&\coloneqq\hat{\vierer{r}}_{\rho}\delta^{\Trans}_{\nu\sigma}+\hat{\vierer{r}}_{\sigma}\delta^{\Trans}_{\nu\rho} &
	F^{(2)}_{\nu\rho\sigma}(\hat{\vierervec{r}})&\coloneqq\hat{\vierer{r}}_{\nu}\delta_{\rho\sigma} &
	F^{(3)}_{\nu\rho\sigma}(\hat{\vierervec{r}})&\coloneqq\hat{\vierer{r}}_{\nu}\Loperator_{\rho\sigma},
\end{align}
which are allowed on symmetry grounds, following the definitions~\cref{eq:TD_zT_L_operator_definition,eq:TD_zT_delta_transverse_definition}.
The calculation of the derivative of the correlation function in its tensor-decomposed form is straightforward and can be expressed via the three upper tensorial structures as
\begin{align}
	\begin{split}
		0&=
		\begin{aligned}[t]
			&\left[
				\left(-\frac{D}{2}+\frac{1}{D-1}\right)G_{\Trans\Trans}
				+G'_{\Longi\Trans}
				+\frac{D}{2}G_{\Longi\Trans}
				-\frac{1}{D-1}G_{\longi\longi}
				+\frac{1}{\sqrt{D-1}}G_{\scal\longi}
			\right]F^{(1)}_{\nu\rho\sigma} \\
			+&\left[
				\frac{2}{D}G'_{\scal\scal}
				+\frac{2\sqrt{D-1}}{D}G'_{\scal\longi}
				+\sqrt{D-1}G_{\scal\longi}
			\right]F^{(2)}_{\nu\rho\sigma} \\
			+&\left[
				-G_{\Longi\Trans}
				+\frac{2}{D}G'_{\longi\longi}
				+G_{\longi\longi}
				+\frac{2}{D\sqrt{D-1}}G'_{\scal\longi}
			\right]F^{(3)}_{\nu\rho\sigma}.
		\end{aligned}
	\end{split}
\end{align}
Here, we introduced a dimensionless definition of the derivative as
\begin{align}
	G'_{X}(\vierer{r}^{2})&\coloneqq \vierer{r}^{2}\left(\dd{}{\vierer{r}^{2}}G_{X}\right)(\vierer{r}^{2}).
\end{align}
Terms without $\vierer{r}^{2}$-derivatives arise from the $\vierervec{r}$ dependence of the projection operators.
As mentioned in the beginning of this section, for the r.h.s.\ to vanish, each coefficient of the $F^{(i)}$ tensors has to vanish resulting in three differential relations
\begin{subequations}\label{eq:EMC_zT_constraints}
	\begin{align}
		(D-1) G'_{\Longi\Trans}
		&=
		\frac{D^{2}{-}D{-}2}{2} G_{\Trans\Trans}
		-\frac{D(D{-}1)}{2} G_{\Longi\Trans}
		+ G_{\longi\longi}
		-\sqrt{D-1} G_{\scal\longi}
		\\
		\frac{2}{\sqrt{D-1}} G'_{\scal\scal}
		+2 G'_{\scal\longi}
		&=
		-D G_{\scal\longi}
		\\
		\frac{2}{\sqrt{D-1}}G'_{\scal\longi}
		+2G'_{\longi\longi}
		&=
		-D G_{\longi\longi}
		+D G_{\Longi\Trans}.
	\end{align}
\end{subequations}
These three interrelations between five functions imply that there are only two truly independent correlations functions after considering EMC.
We will see this explicitly when we find a spectral decomposition of these functions in~\cref{subsec:SD_zT}.

\subsection{Finite Temperature --- Averaging over time} \label{subsec:EMC_fT_aTau}

Similarly to the previous section, the conservation of energy and momentum implies interrelations between the ten correlation functions of EMT correlators.
The procedure is the same as in~\cref{eq:EMT_correlator_conservation_as_Fs}: The full relation of EMC for the correlation function is calculated and expressed in terms of all tensor structures $F^{(i)}_{\nu\rho\sigma}$, which respect all relevant symmetries.
For the full equation to vanish, each coefficient of $F^{(i)}_{\nu\rho\sigma}$ has to vanish resulting in one differential equation each.

Analogous to the tensor decomposition of the EMT correlator in this scenario, the previously found tensors~\eqref{eq:EMC_zT_F_tensors} are extended by two tensors which have an even number of temporal components, resulting in the following five tensors:
\begin{subequations}\label{eq:EMC_fT_aTau_F_tensors}
	\begin{align}
		F^{(1)}_{\nu\rho\sigma}&\coloneqq\hat{r}_{\rho}\delta^{\Trans}_{\nu\sigma}+\hat{r}_{\sigma}\delta^{\Trans}_{\nu\rho} &
		F^{(2)}_{\nu\rho\sigma}&\coloneqq\hat{r}_{\nu}\delta_{\rho\sigma} &
		F^{(3)}_{\nu\rho\sigma}&\coloneqq\hat{r}_{\nu}\left(3\hat{r}_{\rho}\hat{r}_{\sigma}-\delta^{\ispace}_{\rho\sigma}\right)
		\label{eq:EMC_fT_aTau_F_tensors_vacuum} \\
		F^{(4)}_{\nu\rho\sigma}&\coloneqq\hat{r}_{\nu}\left(3u_{\rho}u_{\sigma}-\delta^{\ispace}_{\rho\sigma}\right) &
		F^{(5)}_{\nu\rho\sigma}&\coloneqq u_{\nu}\left(\hat{r}_{\rho}u_{\sigma}+\hat{r}_{\sigma}u_{\rho}\right).
		\label{eq:EMC_fT_aTau_F_tensors_new}
	\end{align}
\end{subequations}
We again differentiate $G_{\mu\nu\rho\sigma}$ from~\cref{eq:TD_fT_aTau_G}, remembering to differentiate both the projectors and the component functions, organize the results in terms of the $F^{(i)}_{\nu\rho\sigma}$, and demand that the prefactor on each $F^{(i)}$ be zero, resulting in five differential equations:
\begin{subequations}\label{eq:EMC_fT_aTau_constraints}
	\begin{align}
		\label{eq:EMC_fT_aTau_constraint_1}
		-G'_{\Longi\Trans}
		&=-G_{\Trans\Trans} +
		\frac{3}{2} G_{\Longi\Trans}
		+ \frac{\sqrt{3} G_{\scal\longi} - \sqrt{2} G_{\longi\longi} - G_{\longi\enth}}{\sqrt{8}} \\
		\label{eq:EMC_fT_aTau_constraint_2}
		\frac{\sqrt{3} G'_{\scal\scal} + \sqrt{8} G'_{\scal\longi} - G'_{\scal\enth}}{\sqrt{18}}
		&=-G_{\scal\longi} \\
		\label{eq:EMC_fT_aTau_constraint_3}
		\frac{\sqrt{3} G'_{\scal\longi} + \sqrt{8} G'_{\longi\longi} - G'_{\longi\enth}}{\sqrt{18}}
		&=-G_{\longi\longi} 
		+G_{\Longi\Trans} \\
		\label{eq:EMC_fT_aTau_constraint_4}
		\frac{\sqrt{3} G'_{\scal\enth} + \sqrt{8} G'_{\longi\enth} - G'_{\enth\enth}}{\sqrt{18}}
		&=-G_{\longi\enth} \\
		\label{eq:EMC_fT_aTau_constraint_5}
		-G'_{\iTime\Longi}
		&=G_{\iTime\Longi}
		-G_{\iTime\Trans}
	\end{align}
\end{subequations}
Some remarks:
\begin{enumerate}
	\item\label{itm:EMC_fT_aTau_remark_1} The two component functions $G_{\iTime\Longi}$ and $G_{\iTime\Trans}$ represent correlators of $T_{0i}$-type EMT components which are odd under time reflection while the other eight component functions correspond to correlation functions of time-even EMT components.
		As the two kinds of EMT components transform differently under time reflection, the five constraints do not mix the two subsets of correlator components.
	\item The three relations~\cref{eq:EMC_fT_aTau_constraint_2,eq:EMC_fT_aTau_constraint_3,eq:EMC_fT_aTau_constraint_4}
		are the same except with the sequential substitution $\scal \to \longi \to \enth$ in one index, except that~\cref{eq:EMC_fT_aTau_constraint_3} has a $G_{\Longi\Trans}$ factor on the r.h.s.
	\item Certain combinations of $G$ factors in the scalar sector appear repeatedly because the underlying $C$-projectors combine to form particularly simple tensor structures.
		In particular, the combination
		\begin{subequations}
			\begin{align}
				\sqrt{3} C^{\scal}_{\alpha\beta} + \sqrt{8} C^{\longi}_{\alpha\beta} -  C^{\enth}_{\alpha\beta}
				&= \sqrt{12} \: \hat{r}_{\alpha} \hat{r}_{\beta}
				\eqqcolon \sqrt{12} \: C^{\sep}_{\alpha\beta}
				\intertext{appears in~\cref{eq:EMC_fT_aTau_constraint_2,eq:EMC_fT_aTau_constraint_3,eq:EMC_fT_aTau_constraint_4} and the combination}
				\sqrt{3} C^{\scal}_{\alpha\beta} - \sqrt{2} C^{\longi}_{\alpha\beta} - C^{\enth}_{\alpha\beta}
				&= \sqrt{3} \: \delta^{\ispace\Trans}_{\alpha\beta}
				\eqqcolon \sqrt{6} \: C^{\trans}_{\alpha\beta}
			\end{align}
		\end{subequations}
		appears in~\cref{eq:EMC_fT_aTau_constraint_1}, because they both represent the relevant operators for the EMC constraints, which are not part of our scalar basis.
		As mentioned in~\cref{subsubsec:TD_zT_correlator_decomposition},~\cref{footnote:scalar_basis_choice}, the chosen scalar basis is best for our applications but not ideal for the constraints implied by EMC.
\end{enumerate}

\subsection{Finite Temperature --- General time}\label{subsec:EMC_fT_gTau}
In contrast to the previous two sections, the implications of EMC in the case of a general temporal separation in the correlation function are more complicated because the resulting relations are partial differential equations.
However, the procedure of deriving them is similar.
The derivative splits up into its temporal and its spatial part, whereas the temporal part only acts on the correlation function $G^{X}(\tau,r^{2})$ while the spatial part \emph{also} acts on the projector $P^{X}_{\mu\nu\rho\sigma}(\hat{\vec{r}})$.
The full equation for EMC can be structured into three main parts
\begin{subequations}
	\begin{align}
		0&=r\partial_{\mu}G_{\mu\nu\rho\sigma}(\tau,\vec{r}) \\
		&=\sum_{X}\left[
			\dot{G}_{X}(\tau,r^{2})P^{X}_{0\nu\rho\sigma}(\hat{\vec{r}})
			+G'_{X}(\tau,r^{2})\cdot2\hat{r}_{i}P^{X}_{i\nu\rho\sigma}(\hat{\vec{r}})
			+G_{X}(\tau,r^{2})(r\partial_{i}P^{X}_{i\nu\rho\sigma})(\hat{\vec{r}})
		\right]
	\end{align}
\end{subequations}
with the definitions for the temporal and spatial derivatives of $G_{X}$
\begin{align}
	\dot{G}_{X}(\tau,r^{2})&\coloneqq\left(r\pdd{}{\tau}G_{X}\right)(\tau,r^{2}) &
	G'_{X}(\tau,r^{2})&\coloneqq\left(r^{2}\pdd{}{r^{2}}G_{X}\right)(\tau,r^{2})
\end{align}
to absorb some factors due to dimensionality.
Aside from calculating the contractions of $\hat{\vec{r}}$ with $P^{X}$ and the spatial derivative of $P^{X}$, one has to evaluate $P^{X}_{0\nu\rho\sigma}$ too, which is also straightforward.

As the set of symmetries is even further reduced compared to the previous two sections, there are now ten possible $3$-index structures:
\begin{align}
	\begin{split}
		\begin{aligned}[c]
			F^{(1)}_{\nu\rho\sigma}&\coloneqq\hat{r}_{\rho}\delta^{\ispace\Trans}_{\nu\sigma}+\hat{r}_{\sigma}\delta^{\ispace\Trans}_{\nu\rho} &
			F^{(2)}_{\nu\rho\sigma}&\coloneqq\hat{r}_{\nu}\delta_{\rho\sigma} &
			F^{(3)}_{\nu\rho\sigma}&\coloneqq\hat{r}_{\nu}\left(3\hat{r}_{\rho}\hat{r}_{\sigma}-\delta^{\ispace}_{\rho\sigma}\right)
			\\
			F^{(4)}_{\nu\rho\sigma}&\coloneqq\hat{r}_{\nu}\left(3u_{\rho}u_{\sigma}-\delta^{\ispace}_{\rho\sigma}\right) &
			F^{(5)}_{\nu\rho\sigma}&\coloneqq u_{\nu}\left(\hat{r}_{\rho}u_{\sigma}+\hat{r}_{\sigma}u_{\rho}\right) &&
			\\
			F^{(6)}_{\nu\rho\sigma}&\coloneqq u_{\rho}\delta^{\ispace\Trans}_{\nu\sigma}+u_{\sigma}\delta^{\ispace\Trans}_{\nu\rho} &
			F^{(7)}_{\nu\rho\sigma}&\coloneqq u_{\nu}\delta_{\rho\sigma} &
			F^{(8)}_{\nu\rho\sigma}&\coloneqq u_{\nu}\left(3\hat{r}_{\rho}\hat{r}_{\sigma}-\delta^{\ispace}_{\rho\sigma}\right)
			\\
			F^{(9)}_{\nu\rho\sigma}&\coloneqq u_{\nu}(3u_{\rho}u_{\sigma}-\delta^{\ispace}_{\rho\sigma}) &
			F^{(10)}_{\nu\rho\sigma}&\coloneqq\hat{r}_{\nu}\left(\hat{r}_{\rho}u_{\sigma}+u_{\rho}\hat{r}_{\sigma}\right) &&
		\end{aligned}
	\end{split}
\end{align}
extending the previously found tensors by five time-odd tensors.

Overall, the total equation for EMC can be expressed in terms of these ten tensors $F^{(i)}_{\nu\rho\sigma}$ resulting in ten partial differential equations
\begin{subequations}
	\label{eq:EMC_fT_gTau_constraints}
	\begin{align}
		\label{eq:EMC_fT_gTau_constraint_1}
		-\frac{1}{2}\dot{G}_{\Mix\Trans}
		-G'_{\Longi\Trans}
		&=
		-G_{\Trans\Trans}
		+ \frac{3}{2} G_{\Longi\Trans}
		+ \frac{\sqrt{3} G_{\scal\longi} - \sqrt{2}G_{\longi\longi} - G_{\longi\enth}}{\sqrt{8}}
		\\
		\label{eq:EMC_fT_gTau_constraint_2}
		\frac{1}{\sqrt{12}}\dot{G}_{\scal\mix}
		+\frac{\sqrt{3}G'_{\scal\scal}+\sqrt{8}G'_{\scal\longi}-G'_{\scal\enth}}{\sqrt{18}}
		&=
		-G_{\scal\longi}
		\\
		\label{eq:EMC_fT_gTau_constraint_3}
		\frac{1}{\sqrt{12}}\dot{G}_{\longi\mix}
		+\frac{\sqrt{3}G'_{\scal\longi}+\sqrt{8}G'_{\longi\longi}-G'_{\longi\enth}}{\sqrt{18}}
		&=
		-G_{\longi\longi}
		+G_{\Longi\Trans}
		\\
		\label{eq:EMC_fT_gTau_constraint_4}
		\frac{1}{\sqrt{12}}\dot{G}_{\enth\mix}
		+\frac{\sqrt{3}G'_{\scal\enth}+\sqrt{8}G'_{\longi\enth}-G'_{\enth\enth}}{\sqrt{18}}
		&=
		-G_{\longi\enth}
		\\
		\label{eq:EMC_fT_gTau_constraint_5}
		\frac{\dot{G}_{\scal\mix}+\sqrt{3}\dot{G}_{\enth\mix}}{\sqrt{8}}
		+G'_{\mix\mix}
		&=
		G_{\iTime\Trans}
		-G_{\mix\mix}
		\\
		\label{eq:EMC_fT_gTau_constraint_6}
		-\frac{1}{2}\dot{G}_{\iTime\Trans}
		-G'_{\Mix\Trans}
		&=
		\frac{3}{2}G_{\Mix\Trans}
		+ \frac{\sqrt{3} G_{\scal\mix} - \sqrt{2} G_{\longi\mix} - G_{\enth\mix}}{\sqrt{24}}
		\\
		\label{eq:EMC_fT_gTau_constraint_7}
		\frac{\dot{G}_{\scal\scal}+\sqrt{3}\dot{G}_{\scal\enth}}{\sqrt{8}}
		+G'_{\scal\mix}
		&=
		-G_{\scal\mix}
		\\
		\label{eq:EMC_fT_gTau_constraint_8}
		\frac{\dot{G}_{\scal\longi}+\sqrt{3}\dot{G}_{\longi\enth}}{\sqrt{8}}
		+G'_{\longi\mix}
		&=
		-G_{\longi\mix}
		+\sqrt{3}G_{\Mix\Trans}
		\\
		\label{eq:EMC_fT_gTau_constraint_9}
		\frac{\dot{G}_{\scal\enth}+\sqrt{3}\dot{G}_{\enth\enth}}{\sqrt{8}}
		+G'_{\enth\mix}
		&=
		-G_{\enth\mix}
		\\
		\label{eq:EMC_fT_gTau_constraint_10}
		\frac{1}{2}\dot{G}_{\mix\mix}
		+\frac{\sqrt{3}G'_{\scal\mix}+\sqrt{8}G'_{\longi\mix}-G'_{\enth\mix}}{\sqrt{6}}
		&=
		G_{\Mix\Trans}
		-\sqrt{3}G_{\longi\mix}
	\end{align}
\end{subequations}

\noindent Again, some remarks:
\begin{enumerate}
	\item The first five constraints,
		\cref{eq:EMC_fT_gTau_constraint_1,eq:EMC_fT_gTau_constraint_2,eq:EMC_fT_gTau_constraint_3,eq:EMC_fT_gTau_constraint_4,eq:EMC_fT_gTau_constraint_5},
		are the same as from the previous~\namecref{subsec:EMC_fT_aTau},
		\cref{eq:EMC_fT_aTau_constraint_1,eq:EMC_fT_aTau_constraint_2,eq:EMC_fT_aTau_constraint_3,eq:EMC_fT_aTau_constraint_4,eq:EMC_fT_aTau_constraint_5},
		only extended by time derivatives of the time-odd component functions.
	\item The other five constraints,
		\cref{eq:EMC_fT_gTau_constraint_6,eq:EMC_fT_gTau_constraint_7,eq:EMC_fT_gTau_constraint_8,eq:EMC_fT_gTau_constraint_9,eq:EMC_fT_gTau_constraint_10},
		are new and interrelate time derivatives of time-even component functions with spatial derivatives of time-odd component functions and with time-odd component functions themselves.
	\item In analogy to the relations from~\cref{subsec:EMC_fT_aTau},
		\cref{eq:EMC_fT_aTau_constraint_2,eq:EMC_fT_aTau_constraint_3,eq:EMC_fT_aTau_constraint_4},
		the relations
		\cref{eq:EMC_fT_gTau_constraint_2,eq:EMC_fT_gTau_constraint_3,eq:EMC_fT_gTau_constraint_4}
		are the same except with the sequential substitutions $\scal \to \longi \to \enth$ in one index, only with the longitudinal relation~\cref{eq:EMC_fT_gTau_constraint_3} connecting to the vector component function $G_{\Longi\Trans}$.
		The same holds true for the three, new relations
		\cref{eq:EMC_fT_gTau_constraint_7,eq:EMC_fT_gTau_constraint_8,eq:EMC_fT_gTau_constraint_9},
		where again the longitudinal relation~\cref{eq:EMC_fT_gTau_constraint_8} connects to the vector component function $G_{\Mix\Trans}$.
	\item Again, in the scalar sector certain combinations of $G$ functions occur because the underlying $C$-functions represent particular simple tensor structures:
		\begin{subequations}
			\begin{align}
				\sqrt{3} C^{\scal}_{\mu\nu} + \sqrt{8} C^{\longi}_{\mu\nu} - C^{\enth}_{\mu\nu} & = \sqrt{12}\:  \hat r_{\mu} \hat r_{\nu}, \\
				C^{\scal}_{\mu\nu} + \sqrt{3} C^{\enth}_{\mu\nu} & = 2\: u_{\mu} u_{\nu}, \\
				\sqrt{3} C^{\scal}_{\mu\nu} - \sqrt{2} C^{\longi}_{\mu\nu} - C^{\enth}_{\mu\nu} & = \sqrt{3}\: \delta^{\ispace\Trans}_{\mu\nu}.
			\end{align}
		\end{subequations}
\end{enumerate}

\noindent One way to try to understand these PDEs better is to Fourier transform the time direction into Matsubara frequencies $\omega_{n}=2\pi nT$, or equivalently, to write
\begin{subequations}
	\begin{align}
		G^{X}(\tau,r^{2}) & = G^{X}_{0}(r^{2}) + \sum_{n=1}^{\infty} G^{X}_{n}(r^{2}) \cos(\omega_{n}\tau)
		\,, & X & \in \mathcal{I}_{\bar{\tau}}
		\\
		G^{Y}(\tau,r^{2}) & = \sum_{n=1}^{\infty} G^{Y}_n(r^{2}) \sin(\omega_{n}\tau) \,, &
		Y & \in \mathcal{I}_{\tau} \setminus \mathcal{I}_{\bar{\tau}}
	\end{align}
\end{subequations}
Integrating over $\tau$, the first five equations become~\cref{eq:EMC_fT_aTau_constraints} and the next five vanish.
This case reduces precisely to the previous subsection.
Integrating $\int_{0}^{\beta} \cos(\omega_{n}\tau) \d{\tau}$, the first five equations have $G^X(\tau,r^{2}) \mapsto G^X_{n}(r^{2})$ and $\dot{G}^Y(\tau,r^{2}) \mapsto \omega_{n} r G^Y_{n}(r^{2})$;
integrating $\int_{0}^{\beta} \sin(\omega_{n}\tau) \d{\tau}$, the lower 5 equations have $G^Y(\tau,r^{2}) \mapsto G^Y_{n}(r^{2})$ and $\dot{G}^X(\tau,r^{2}) \mapsto -\omega_{n} r G^X_{n}(r^{2})$.
Therefore, within each non-zero sector, the equations become 10 ODEs in 14 variables, meaning that there are 4 independent functions in each $n$-sector.
The reason that there are 4 functions at nonzero $n$ but 5 at $n=0$ is not entirely clear to us.

\section{Spectral Decomposition}\label{sec:SD}

The constraints we found in the previous section imply that it should be possible to write our $G^{X}$ functions in terms of a smaller number of functions.
This section will partially realize this by writing a spectral decomposition of the component functions in terms of a set of spectral functions, representing the true number of independent variables.
However, we will only be able to achieve this for the vacuum case and the time-averaged thermal case, where the relations are ordinary and not partial differential equations.

\subsection{Zero Temperature} \label{subsec:SD_zT}
To clarify the meaning of the relation between correlation functions implied by EMC, we consider the averaged correlation function in four dimensions
\begin{align}
	A_{\mu\nu\rho\sigma}(\tau)&\coloneqq\int_{-\infty}^{\infty}\d*[D-1]{x_{\perp}}\ev{T_{\mu\nu}(0,x,y,z)T_{\rho\sigma}(-\tau,0,0,0)},
\end{align}
where the separation is in the temporal direction $\vierervec{r}=\tau\vierervec{e}_{0}$
and the left operator is averaged over the hyperplane transversal to the separation.
For $\mu=0$, the integral over $T_{0\nu}(x,y,z,t=0)$ results in the momentum operator $P_{\nu}(t=0)$ and the upper integral can be expressed as a quantum mechanical evolution amplitude
\begin{align}
	A_{0\nu\rho\sigma}(\tau)
	&=\bramidket{0}{P_{\nu}(0) \: \e^{-H\tau}\: T_{\rho\sigma}(\vierervec{0})}{0}.
\end{align}
The momentum operator acting on the vacuum vanishes and hence $A_{0\nu\rho\sigma}(\tau)=0$.
By exchange symmetry, all four indices must be spatial.
But $A_{ijkl}$ is the correlator of two spatial rank-2 tensors, which can be decomposed into a traceless-tensor and a pure-trace part.
These can be expressed via states $\ket{\pi}\coloneqq\sum_{k}T_{kk}\ket{0}$ and $\ket{\pi_{ij}}\coloneqq (T_{ij}-\delta_{ij}\sfrac{T_{kk}}{3})\ket{0}$ as
\begin{align}
	A_{ijkl}(\tau)
	&\coloneqq\Bramidket{0}{T_{ij}(\vec{0})\e^{-H\tau}\int\d*[3]{r}T_{kl}(\vec{r})}{0} \\
	&=\left(\bra{\pi_{ij}}+\frac{\delta_{ij}}{3}\bra{\pi}\right)\e^{-H\tau}\left(\ket{\pi_{kl}}+\frac{\delta_{kl}}{3}\ket{\pi}\right).
\end{align}
As these states are linear combinations of zero-momentum single- and multiple-particle states, which arise at a range of masses, they can be expressed in terms of two spectral functions in a Lehmann-type representation via
\begin{align}
	A_{ijkl}(\tau)&=\int_{0}^{\infty}\d*{m}m\e^{-m\tau}\left[
		\rho^{\Trans\Trans}(m)P^{\Trans\Trans}_{ijkl}(\vierervec{u})
		+\rho^{\trans\trans}(m)P^{\trans\trans}_{ijkl}(\vierervec{u})
	\right]
\end{align}
with the doubly-transversal, tensorial projector $P^{\Trans\Trans}$ from before~\eqref{eq:TD_fT_aTau_P_TT} and the doubly-transversal, scalar projector $P^{\trans\trans}$
\begin{align}
	P^{\Trans\Trans}_{\mu\nu\rho\sigma}&=\frac{1}{2}(
		\delta^{\Trans}_{\mu\rho}\delta^{\Trans}_{\nu\sigma}
		+\delta^{\Trans}_{\mu\sigma}\delta^{\Trans}_{\nu\rho}
	)
	-\frac{1}{D-1}\delta^{\Trans}_{\mu\nu}\delta^{\Trans}_{\rho\sigma}
	&
	P^{\trans\trans}_{\mu\nu\rho\sigma}&=\frac{1}{D-1}\delta^{\Trans}_{\mu\nu}\delta^{\Trans}_{\rho\sigma}.
\end{align}
Here $\delta^{\Trans}_{\mu\nu}$ is transverse with respect to $\vierer{u}^\mu$, i.e.\ $\delta^{\Trans}_{\mu\nu} = \delta_{\mu\nu} - \vierer{u}_\mu \vierer{u}_\nu$.

Fourier transforming the time direction, and noting that $\vierervec{u}=\hat{\vierervec{p}}$ because we are at vanishing spatial momentum, the expression becomes
\begin{align}\label{eq:SD_zT_Gp_via_rho_and_P}
	G_{\mu\nu\rho\sigma}(\vierervec{p})
	&=\int_{0}^{\infty}\d*{m}\frac{m}{\vierer{p}^{2}+m^{2}}
	\sum_{S\in\mathcal{S}_{0}}\rho^{S}(m)P^{S}_{\mu\nu\rho\sigma}(\hat{\vierervec{p}})
\end{align}
with $S\in\mathcal{S}_{0}=\set{\Trans\Trans,\trans\trans}$ indexing the two spectral functions for the zero temperature case and their projectors.
This expression was derived for $\vierervec{p}\propto\vierervec{u}$, but as written it is covariant and it therefore holds for general $\vierervec{p}$.
In other words, $G_{\mu\nu\rho\sigma}(\vierervec{p})$ is determined by the same tensor decomposition, in $\vierervec{p}$-space, as $G_{\mu\nu\rho\sigma}(\vierervec{r})$ is in coordinate space.
But EMC forbids any $\vierervec{p}$-space component parallel to the momentum $\vierervec{p}$, which excludes $G_{\Longi\Trans}$ and two linear combinations of $C^{\scal}_{\mu\nu}$ and $C^{\longi}_{\mu\nu}$, see~\cref{eq:EMC_momentum_space}, leaving only the two spectral functions shown above.

To see what this implies for $G_{\mu\nu\rho\sigma}(\vierervec{r})$, we Fourier transform to coordinate space:
\begin{align}
	G_{\mu\nu\rho\sigma}(\vierervec{r}) &= \int\fracd*[D]{\vierer{p}}{(2\pi)}\e^{\i\vierervec{p}\cdot\vierervec{r}}
	G_{\mu\nu\rho\sigma}(\vierervec{p}).
\end{align}
Applying the $P^{X}$ projector to extract our individual correlation functions, we find
\begin{align}
	\label{eq:SD_zT_GX_via_rho_and_PP}
	G^{X}(\vierer{r}^{2})&=\int\fracd*[D]{\vierer{p}}{(2\pi)} \e^{\i\vierervec{p}\cdot\vierervec{r}}
	\int_{0}^{\infty}\d*{m}\frac{m}{\vierer{p}^{2}+m^{2}}\sum_{S\in\mathcal{S}_{0}}
	\rho^{S}(m)\frac{1}{M^{X}}P^{S}_{\mu\nu\rho\sigma}(\hat{\vierervec{p}})P^{X}_{\mu\nu\rho\sigma}(\hat{\vierervec{r}})
\end{align}
with $M^{X}$ denoting the multiplicity of each projector from before.
Here, the arguments of both appearing projectors are not aligned, because the integration over $\vierervec{p}$ covers all directions, not only those parallel to $\vierervec{r}$, leading to non-trivial $\hat{\vierervec{p}}$-dependent terms which appear in the integral when contracting the projectors.
Carrying out the contraction of $P^{S}(\hat{\vierervec{p}})$ with $P^{X}(\hat{\vierervec{r}})$ results in a polynomial of the monomials $\set{1,\hat{\vierer{p}}_{\vierer{r}}^{2},\hat{\vierer{p}}_{\vierer{r}}^{4}}$ with $\hat{\vierer{p}}_{\vierer{r}}\coloneqq\hat{\vierervec{p}}\cdot\hat{\vierervec{r}}$.
The coefficients $c^{S,X}_{1,2,3}$ are the respective coefficients to the three monomials $\set{1,\hat{\vierer{p}}_{\vierer{r}}^{2},\hat{\vierer{p}}_{\vierer{r}}^{4}}$ in the contraction of the two projectors, i.e.
\begin{align}
	\frac{1}{M^{X}}P^{S}_{\mu\nu\rho\sigma}(\hat{\vierervec{p}})P^{X}_{\mu\nu\rho\sigma}(\hat{\vierervec{r}})
	&=c^{S,X}_{1}+c^{S,X}_{2}\hat{\vierer{p}}_{\vierer{r}}^{2}+c^{S,X}_{3}\hat{\vierer{p}}_{\vierer{r}}^{4},
\end{align}
which are displayed in~\cref{tab:SD_zT_coefficients} below.
The upper expression~\eqref{eq:SD_zT_GX_via_rho_and_PP} thereby reduces to
\begin{align}\label{eq:SD_zT_GX_via_rho_and_c}
	G^{X}(\vierer{r}^{2})
	&=\int\fracd*[D]{\vierer{p}}{(2\pi)}\e^{\i\vierervec{p}\cdot\vierervec{r}}
	\int_{0}^{\infty}\d*{m}\frac{m}{\vierer{p}^{2}+m^{2}}
	\sum_{S\in\mathcal{S}_{0}}\rho^{S}(m)\big(
		c^{S,X}_{1}
		+c^{S,X}_{2}\hat{\vierer{p}}_{\vierer{r}}^{2}
		+c^{S,X}_{3}\hat{\vierer{p}}_{\vierer{r}}^{4}
	\big).
\end{align}

This integral can be carried out by first performing the Fourier integral, i.e.\
\begin{align}
	G^{X}(\vierer{r}^{2})&=\int_{0}^{\infty}\d*{m}m\sum_{S\in\mathcal{S}_{0}}\rho^{S}(m)\left[
		\int\fracd*[D]{\vierer{p}}{(2\pi)}\frac{\e^{\i\vierervec{p}\cdot\vierervec{r}}}{\vierer{p}^{2}+m^{2}}\big(
			c^{S,X}_{1}
			+c^{S,X}_{2}\hat{\vierer{p}}_{\vierer{r}}^{2}
			+c^{S,X}_{3}\hat{\vierer{p}}_{\vierer{r}}^{4}
		\big)
	\right],
\end{align}
which evaluates to
\begin{align}
	\label{eq:SD_zT_GX_closed_form}
	G^{X}(\vierer{r}^{2})&=\int_{0}^{\infty}\d*{m}m
	\sum_{S\in\mathcal{S}_{0}}\rho^{S}(m)\left[
		c^{S,X}_{1}f_{1}(m,\vierer{r})
		+c^{S,X}_{2}f_{2}(m,\vierer{r})
		+c^{S,X}_{3}f_{3}(m,\vierer{r})
	\right].
\end{align}
Here, the three functions $f_{1,2,3}$ appear, which are defined by
\begin{align}
	f_{1}(m,\vierer{r})&\coloneqq\int\fracd*[D]{\vierer{p}}{(2\pi)}\frac{\e^{\i\vierervec{p}\cdot\vierervec{r}}}{\vierer{p}^{2}+m^{2}} &
	f_{n+1}(m,\vierer{r})&\coloneqq\frac{1}{m^{2}}\dd[2]{}{\vierer{r}}f_{n}(m,\vierer{r}) \, , \quad n\in\mathds{N}
\end{align}
and evaluate for $D=4$ and $D=3$ to
\begin{subequations}
	\label{eq:SD_zT_functions}
	\begin{align}
		f_{1}(m,\vierer{r)}
		&=
		\begin{dcases}
			\frac{m^{2}}{4\pi^{2}}\frac{K_{1}(m\vierer{r})}{m\vierer{r}} & ,\ D=4 \\
			\frac{m}{4\pi}\frac{\e^{-m\vierer{r}}}{m\vierer{r}} & ,\ D=3
		\end{dcases}
		\\
		f_{2}(m,\vierer{r})
		&=
		\begin{dcases}
			\frac{m^{2}}{4\pi^{2}(m\vierer{r})^{3}}\left(
				3m\vierer{r} K_{0}(m\vierer{r})
				+\left((m\vierer{r})^{2}+6\right) K_{1}(m\vierer{r})
			\right) & ,\ D=4 \\
			\frac{m \e^{-m\vierer{r}}}{4\pi (m\vierer{r})^{3}} \left( (m\vierer{r})^{2}+2m\vierer{r}+2 \right)
			& ,\ D=3
		\end{dcases} \\
		f_{3}(m,r)
		&=
		\begin{dcases}
			\begin{aligned}
				\frac{m^{2}}{4\pi^{2}(m\vierer{r})^{5}}\big(
					&(6(m\vierer{r})^{3}+60m\vierer{r})K_{0}(m\vierer{r}) \\
					&+\left( (m\vierer{r})^{4} + 27 (m\vierer{r})^{2} + 120 \right) K_{1}(m\vierer{r})
				\big)
			\end{aligned} & ,\ D=4 \\
			\frac{m \e^{-m\vierer{r}}}{4\pi (m\vierer{r})^{5}}
			\left( (m\vierer{r})^{4} + 4 (m\vierer{r})^{3} + 12 (m\vierer{r})^{2} + 24 m\vierer{r} + 24 \right)
			& ,\ D=3
		\end{dcases},
	\end{align}
\end{subequations}
where $K_{0,1}$ is the modified Bessel function of the second kind.
We have checked that \eqref{eq:SD_zT_GX_closed_form} for general non-negative spectral functions generates component functions which satisfy the constraints, \eqref{eq:EMC_zT_constraints}.
Since \eqref{eq:EMC_zT_constraints} represent three constraints on five functions, leaving two independent functions, this proves that \eqref{eq:SD_zT_GX_closed_form} gives the most general form for the component functions.

\begin{table}[!ht]
	\centering
	\begin{subtable}[c]{\textwidth}
		\centering
		\begin{tabular}{@{} c*{6}{c} @{}}
			\toprule
			$X$ & $c^{\Trans\Trans,X}_{1}$ & $c^{\Trans\Trans,X}_{2}$ & $c^{\Trans\Trans,X}_{3}$ & $c^{\trans\trans,X}_{1}$ & $c^{\trans\trans,X}_{2}$ & $c^{\trans\trans,X}_{3}$ \\
			\midrule
			$\Trans\Trans$ & $\!\frac{D^{3}{-}3D^{2}+D{-}1}{(D{+}1)(D{-}1)^{2}}\!$ & $\!\frac{2(D^{2}{-}2D{+}3)}{(D{+}1)(D{-}1)^{2}}\!$ & $\!\frac{2(D{-}2)}{(D{+}1)(D{-}1)^{2}}\!$ & $\!\frac{2}{(D{+}1)(D{-}1)^{2}}\!$ & $-\frac{4}{(D{+}1)(D{-}1)^{2}}\!$ & $\!\frac{2}{(D{+}1)(D{-}1)^{2}}\!$ \\\addlinespace
			$\Longi\Trans$ & $\frac{D{-}2}{D{-}1}$ & $-\frac{(D{-}2)(D{-}3)}{(D{-}1)^{2}}$ & $-\frac{2(D{-}2)}{(D{-}1)^{2}}$ & $0$ & $\frac{2}{(D{-}1)^{2}}$ & $-\frac{2}{(D{-}1)^{2}}$ \\\addlinespace
			$\longi\longi$ & $\frac{D(D{-}2)}{(D{-}1)^{2}}$ & $-\frac{2D(D{-}2)}{(D{-}1)^{2}}$ & $\frac{D(D{-}2)}{(D{-}1)^{2}}$ & $\frac{1}{D(D{-}1)^{2}}$ & $-\frac{2}{(D{-}1)^{2}}$ & $\frac{D}{(D{-}1)^{2}}$ \\\addlinespace
			$\scal\scal$ & $0$ & $0$ & $0$ & $\frac{D{-}1}{D}$ & $0$ & $0$ \\\addlinespace
			$\scal\longi$ & $0$ & $0$ & $0$ & $\frac{1}{D\sqrt{D{-}1}}$ & $-\frac{1}{\sqrt{D{-}1}}$ & $0$ \\
			\bottomrule
		\end{tabular}
		\caption{$D$ Dimensions}
	\end{subtable}

	\medskip

	\begin{subtable}[c]{0.49\textwidth}
		\centering
		\begin{tabular}{@{} c*{6}{c} @{}}
			\toprule
			& \multicolumn{6}{c}{$c^{S,X}_{1}$, $c^{S,X}_{2}$, $c^{S,X}_{3}$}
			\\
			\cmidrule{2-7}
			$X$ &
			\multicolumn{3}{c}{$S=\Trans\Trans$} &
			\multicolumn{3}{c}{$S=\trans\trans$}
			\\
			\midrule
			$\Trans\Trans$ &
			$\frac{19}{45}$ & $\frac{22}{45}$ & $\frac{4}{45}$ &
			$\frac{2}{45}$ & $-\frac{4}{45}$ & $\frac{2}{45}$
			\\\addlinespace
			$\Longi\Trans$ &
			$\frac{2}{3}$ & $-\frac{2}{9}$ & $-\frac{4}{9}$ &
			$0$ & $\frac{2}{9}$ & $-\frac{2}{9}$
			\\\addlinespace
			$\longi\longi$ &
			$\frac{8}{9}$ & $-\frac{16}{9}$ & $\frac{8}{9}$ &
			$\frac{1}{36}$ & $-\frac{2}{9}$ & $\frac{4}{9}$
			\\\addlinespace
			$\scal\scal$ &
			$0$ & $0$ & $0$ &
			$\frac{3}{4}$ & $0$ & $0$
			\\\addlinespace
			$\scal\longi$ &
			$0$ & $0$ & $0$ &
			$\frac{1}{4\sqrt{3}}$ & $-\frac{1}{\sqrt{3}}$ & $0$
			\\
			\bottomrule
		\end{tabular}
		\caption{$D=4$}
	\end{subtable}
	\begin{subtable}[c]{0.49\textwidth}
		\centering
		\begin{tabular}{@{} c*{6}{c} @{}}
			\toprule
			& \multicolumn{6}{c}{$c^{S,X}_{1}$, $c^{S,X}_{2}$, $c^{S,X}_{3}$}
			\\
			\cmidrule{2-7}
			$X$ &
			\multicolumn{3}{c}{$S=\Trans\Trans$} &
			\multicolumn{3}{c}{$S=\trans\trans$}
			\\
			\midrule
			$\Trans\Trans$ &
			$\frac{1}{8}$ & $\frac{3}{4}$ & $\frac{1}{8}$ &
			$\frac{1}{8}$ & $-\frac{1}{4}$ & $\frac{1}{8}$
			\\\addlinespace
			$\Longi\Trans$ &
			$\frac{1}{2}$ & $0$ & $-\frac{1}{2}$ &
			$0$ & $\frac{1}{2}$ & $-\frac{1}{2}$
			\\\addlinespace
			$\longi\longi$ &
			$\frac{3}{4}$ & $-\frac{3}{2}$ & $\frac{3}{4}$ &
			$\frac{1}{12}$ & $-\frac{1}{2}$ & $\frac{3}{4}$
			\\\addlinespace
			$\scal\scal$ &
			$0$ & $0$ & $0$ &
			$\frac{2}{3}$ & $0$ & $0$
			\\\addlinespace
			$\scal\longi$ &
			$0$ & $0$ & $0$ &
			$\frac{1}{3\sqrt{2}}$ & $-\frac{1}{\sqrt{2}}$ & $0$
			\\
			\bottomrule
		\end{tabular}
		\caption{$D=3$}
	\end{subtable}
	\caption{The coefficients $c^{S,A}_{i}$ with $S\in\mathcal{S}_{0}$ and $X\in\mathcal{I}_{0}$ for the functions $f_{i}(m,r)$ with $i\in\set{1,2,3}$ in~\cref{eq:SD_zT_GX_via_rho_and_c}.
	}	\label{tab:SD_zT_coefficients}
\end{table}

The large-distance behavior of the correlation function is controlled by the spectral function at the smallest $m$-value where it is nonzero.
In many cases we have some analytical understanding of the smallest-$m$ part of the spectral function.
For instance, in pure-glue QCD, we know that the $0^{++}$ and $2^{++}$ symmetry channels, which are relevant for $\rho^{\trans\trans}$ and $\rho^{\Trans\Trans}$ respectively, contain single-particle (glueball) states with known masses
\cite{Athenodorou:2020ani} which are well isolated from the next state.
In each case, the large-distance behavior can then be found by replacing $\rho(m)$ with a delta function at the glueball mass, times an unknown coefficient.
This provides a simple, analytically known fitting function for the tails of all 5 component functions in terms of four coefficients:  the two glueball masses and the coefficients on the spectral-function delta functions.
In contrast, for full QCD, the lightest states available in both the $0^{++}$ and the $2^{++}$ channels are two-pion states, and one would have to perform a fit to the beginning of a two-particle cut.

\subsection{Finite Temperature --- Averaging over time} \label{subsec:SD_fT_aTau}
In~\cref{subsec:EMC_fT_aTau}, we showed that the ten component functions of the EMT correlator from~\cref{subsec:TD_fT_aTau} are restricted by five ODEs.
Building onto the results from the previous section, one should be able to describe these ten component functions in terms of five spectral functions.

The arguments below are the same as in the previous subsection.
When considering the e.g.\ $z$-direction as the analytically continued, Minkowskian time direction in a space with a compact, spatial $\tau$-direction, the integral $\int_{0}^{\beta}\d{\tau}\int\d{x}\d{y}\: T_{\mu z}$ gives a conserved charge.
Acting on the vacuum, it gives zero and hence one can rule out correlation functions of these combinations.
Therefore, when integrating over the hyperplane orthogonal to the $z$-axis and considering $z$-dependent correlation functions, only $T_{\mu\nu}$ with $\mu\neq z$ and $\nu\neq z$ survive.
These decompose into the following:
\begin{itemize}[noitemsep]
	\item A spatial, traceless tensor with $2$ independent components: $T_{xy}$, $T_{xx}-T_{yy}$
	\item A spatial, time-odd vector with $2$ independent components: $T_{tx}$, $T_{ty}$
	\item A temporal scalar: $T_{tt}$
	\item A spatial scalar: $T_{xx}+T_{yy}$
\end{itemize}
One expects thereby five independent correlation functions: one for the tensor sector, one time-odd for the vector sector, and three (in the form of a symmetric $2\times 2$ matrix of scalar correlation functions) for the scalar sector.

In other words, in momentum space we can write $G_{\mu\nu\rho\sigma}(\vec{p})$ in terms of projectors $P^{X}_{\mu\nu\rho\sigma}(\hat{\vec{p}})$ which have zero projection against any of ($p_{\mu},p_{\nu},p_{\sigma},p_{\rho}$).
This permits $P^{\Trans\Trans}(\hat{\vec{p}})$ and $P^{\iTime\Trans}(\hat{\vec{p}})$ from~\eqref{eq:TD_fT_aTau_P_TT} and~\eqref{eq:TD_fT_aTau_P_UT}, and a $2\times 2$ matrix in the scalar sector built out of the following scalar rank-2 projectors
\begin{align}
	C^{\trans}_{\mu\nu}(\hat{\vec{p}})&\coloneqq\frac{1}{\sqrt{2}}\left( \delta_{\mu\nu}-u_{\mu} u_{\nu} - \hat{p}_{\mu} \hat{p}_{\nu} \right), &
	C^{\itime}_{\mu\nu}&\coloneqq u_{\mu}u_{\nu},
\end{align}
which themselves are linear combinations of $C^{\scal},C^{\longi},C^{\enth}$.
Here, we choose the $(\trans,\itime)$-basis for the spectral projectors as they are the physically relevant ones.
Associated to these, there are three spectral functions $\rho_{\trans\trans}(m),\rho_{\trans\itime}(m),\rho_{\itime\itime}(m)$, together with $\rho_{\Trans\Trans}(m)$ and $\rho_{\iTime\Trans}(m)$.
In analogy to \cref{eq:SD_zT_Gp_via_rho_and_P}, the momentum-space correlation function is then given in terms of these five spectral functions as:
\begin{align}\label{eq:SD_fT_aTau_Gp_via_rho_and_P}
	G_{\mu\nu\rho\sigma}(\vec{p})
	&=\int_{0}^{\infty}\d*{m}\frac{m}{p^{2}+m^{2}}
	\sum_{S\in\mathcal{S}_{\bar{\tau}}}\rho^{S}(m)P^{S}_{\mu\nu\rho\sigma}(\hat{p})
\end{align}
with $S\in\mathcal{S}_{\bar{\tau}}=\set{\Trans\Trans,\trans\trans,\itime\itime,\trans\itime,\iTime\Trans}$.

Positivity of the spectrum requires
\begin{align}
	\rho^{\trans\trans}\rho^{\itime\itime} &\geq (\rho^{\trans\itime})^{2} &
	\rho^{\trans\trans} &\geq 0 &
	\rho^{\itime\itime} &\geq 0 &
	\rho^{\Trans\Trans} & \geq 0 &
	\rho^{\iTime\Trans} & \geq 0
\end{align}
at every $m$.
To determine the form of the component functions in terms of these spectral functions, we follow the same procedure as in~\cref{eq:SD_zT_GX_via_rho_and_PP} through~\cref{eq:SD_zT_GX_closed_form}, arriving at
\begin{align}
	\label{eq:SD_fT_aTau_GC_closed_form}
	G^{X}(r^{2}) &= \int_{0}^{\infty}\d*{m} m \sum_{S\in\mathcal{S}_{\bar{\tau}}}
	\rho^{S}(m) \left[ c_{1}^{S,X} f_{1}(m,r) + c_{2}^{S,X} f_{2}(m,r) + c_{3}^{S,X} f_{3}(m,r) \right].
\end{align}
The 3D form of the functions $f_{1,2,3}(m,r)$ from~\cref{eq:SD_zT_functions} should be used here, and the coefficients are tabulated in~\cref{eq:SD_fT_aTau_coefficients}.

\begin{table}[htb]
	\centering
	\begin{tabular}{@{} c*{15}{c} @{}}
		\toprule
		& \multicolumn{15}{c}{$c^{S,X}_{1}$, $c^{S,X}_{2}$, $c^{S,X}_{3}$}
		\\
		\cmidrule{2-16}
		$X$ &
		\multicolumn{3}{c}{$S=\Trans\Trans$} &
		\multicolumn{3}{c}{$S=\trans\trans$} &
		\multicolumn{3}{c}{$S=\itime\itime$} &
		\multicolumn{3}{c}{$S=\trans\itime$} &
		\multicolumn{3}{c}{$S=\iTime\Trans$}
		\\
		\midrule
		$\Trans\Trans$ &
		$\frac{1}{8}$ & $\frac{3}{4}$ & $\frac{1}{8}$ &
		$\frac{1}{8}$ & $-\frac{1}{4}$ & $\frac{1}{8}$ &
		0 & 0 & 0 &
		0 & 0 & 0 &
		0 & 0 & 0
		\\\addlinespace
		$\Longi\Trans$ &
		$\frac{1}{2}$ & 0 & $-\frac{1}{2}$ &
		$0$ & $\frac{1}{2}$ & $-\frac{1}{2}$ &
		0 & 0 & 0 &
		0 & 0 & 0 &
		0 & 0 & 0
		\\\addlinespace
		$\longi\longi$ &
		$\frac{3}{4}$ & $-\frac{3}{2}$ & $\frac{3}{4}$ &
		$\frac{1}{12}$ & $-\frac{1}{2}$ & $\frac{3}{4}$ &
		0 & 0 & 0 &
		0 & 0 & 0 &
		0 & 0 & 0
		\\\addlinespace
		$\scal\scal$ &
		0 & 0 & 0 &
		$\frac{1}{2}$ & 0 & 0 &
		$\frac{1}{4}$ & 0 & 0 &
		$\frac{1}{\sqrt{2}}$ & 0 & 0 &
		0 & 0 & 0
		\\\addlinespace
		$\enth\enth$ &
		0 & 0 & 0 &
		$\frac{1}{6}$ & 0 & 0 &
		$\frac{3}{4}$ & 0 & 0 &
		$-\frac{1}{\sqrt{2}}$ & 0 & 0 &
		0 & 0 & 0
		\\\addlinespace
		$\scal\longi$ &
		0 & 0 & 0 &
		$\frac{1}{2\sqrt{6}}$ & $-\frac{\sqrt{3}}{2\sqrt{2}}$ & 0 &
		0 & 0 & 0 &
		$\frac{1}{4\sqrt{3}}$ & $-\frac{\sqrt{3}}{4}$ & 0 &
		0 & 0 & 0
		\\\addlinespace
		$\scal\enth$ &
		0 & 0 & 0 &
		$-\frac{1}{2\sqrt 3}$ & 0 & 0 &
		$\frac{\sqrt 3}{4}$  & 0 & 0 &
		$\frac{1}{\sqrt 6}$  & 0 & 0 &
		0 & 0 & 0
		\\\addlinespace
		$\longi\enth$ &
		0 & 0 & 0 &
		$-\frac{1}{6\sqrt{2}}$ &$\frac{1}{2\sqrt{2}}$ & 0 &
		0 & 0 & 0 &
		$\frac{1}{4}$ & $-\frac{3}{4}$ & 0 &
		0 & 0 & 0
		\\\addlinespace
		$\iTime\Trans$ &
		0 & 0 & 0 &
		0 & 0 & 0 &
		0 & 0 & 0 &
		0 & 0 & 0 &
		$\frac{1}{2}$ & $\frac{1}{2}$ & 0
		\\\addlinespace
		$\iTime\Longi$ &
		0 & 0 & 0 &
		0 & 0 & 0 &
		0 & 0 & 0 &
		0 & 0 & 0 &
		$\frac{1}{2}$ & $-\frac{1}{2}$ & 0
		\\\addlinespace
		\bottomrule
	\end{tabular}
	\caption{
		The coefficients $c^{S,X}_{i}$ with $S\in\mathcal{S}_{\bar{\tau}}$ and $X\in\mathcal{I}_{\bar{\tau}}$ for the functions $f_{i}(m,r)$ with $i\in\set{1,2,3}$ for~\cref{eq:SD_fT_aTau_GC_closed_form}.
	}
	\label{eq:SD_fT_aTau_coefficients}
\end{table}

Because $P^{\iTime\Trans},P^{\iTime\Longi}$ are odd under $\tau$-reflection and the other projectors are even, we find that the coefficient table is block-diagonal, with $\set{ G^{\iTime\Trans},G^{\iTime\Longi} }$ determined by $\rho^{\iTime\Trans}$ and the other component functions determined by the other four spectral functions.

\subsection{Finite Temperature --- General time} \label{subsec:SD_fT_gTau}
As seen in the last EMC~\cref{subsec:EMC_fT_gTau}, we find ten PDEs~\eqref{eq:EMC_fT_gTau_constraints} inter-relating the 14 general component functions.
The techniques of the previous subsections fail in this case, and it is not clear to us how to write the correlation functions in terms of a simpler spectral decomposition.

One approach is to decompose the fields into Matsubara modes by Fourier transforming along the $\tau$ axis.
A major application of the previous subsections will be to inform the fitting of the large-distance behavior of correlation functions, which will be dominated by the $n=0$ and possibly the $n=1$ Matsubara modes.
Computing mode-by-mode, the PDEs are rendered ODEs.
Considering the zeroth Matsubara mode, the time-averaged scenario is recovered and there are five independent degrees of freedom.
For each non-zero Matsubara mode, however, the constraint PDEs transform into ten linearly independent ODEs on the fourteen component functions, resulting only four independent degrees of freedom.
This discrepancy in the number of degrees of freedom is counterintuitive and unclear to us.
We leave the elucidation of this issue for future work.

\section{Applications} \label{sec:app}
A main goal after this work is a more precise calculation of several transport coefficients, including the shear and bulk viscosities, as well as several \enquote{thermodynamic} transport coefficients~\cite{Kovtun:2018dvd}, from lattice QCD.
Here we will show how to adopt the results of the previous sections into a technique for a more efficient evaluation of EMT correlation functions.

The approach we will advocate is the following:
\begin{enumerate}[noitemsep]
	\item First, evaluate all components of $G_{\mu\nu\rho\sigma}(\tau,\vec{r})$ as a function of lattice separation.
		This can be done efficiently by (fast) Fourier transforming $T_{\mu\nu}(t,\vec{x})$ evaluated at every lattice site $(t,\vec{x})$, considering all products $T_{\mu\nu}(\omega,\vec{k}) T_{\rho\sigma}(-\omega,-\vec{k})$ in momentum space, and Fourier transforming back~\cite{Liu:2017man}.
		In the case that one is interested in $\tau$-integrated correlation functions, only the spatial inverse Fourier transform needs to be taken.
	\item For every separation $\vec{r}$, $G_{\mu\nu\rho\sigma}$ is then decomposed into its 10 (for $\tau$-averaged data) or 14 (for $(\tau,\vec{r})$ dependent data) component functions using \eqref{eq:fT_GXfromPG} or \eqref{eq:aTauGXfromPG}.
	\item The component functions are averaged over distinct spatial $\vec{r}$ values with the same $r^{2}$ value, giving $G^X(\tau,r^2)$ at a discrete set of $r^2$ values.%
		\footnote{
			When projecting lattice data to a functional form which assumes $\text{SO}(3)$ invariance and when averaging over lattice sites with the same $r^{2}$ value but which differ under the discrete lattice symmetries, one loses rotation-non-invariant lattice-artifact effects from the data.
			This is acceptable, because these are not desired anyway and because they should disappear if one performs a continuum extrapolation.
		}
	\item Continuum integrals $4\pi\int\d{r}r^{2}\, G(\tau,r^2)$ or $2\pi^2 \int\d{r} r^3\, G(r^2)$ in the 4D vacuum, become sums $a^D \sum_{n=0}^\infty d_n G(\tau,na^2)$ with $a$ the lattice spacing and $d_n$ the multiplicity with which $r^2=na^2$ occurs, which is OEIS series \texttt{A000118} in 4D and OEIS series \texttt{A005875} in 3D\footnote{
		\href{https://oeis.org/A000118}{https://oeis.org/A000118} and \href{https://oeis.org/A005875}{https://oeis.org/A005875} respectively.}.
	\item\label{itm:app_last_step} Finally, one must convert the component functions $G^{X}(r^{2})$ into information about the transport coefficients of interest.
\end{enumerate}
Reducing the lattice data to a handful of component functions, which depend only on $r^{2}$, represents an extremely efficient, but in-principle lossless, reduction of the correlation-function data.
This is a principal advantage of our approach.
But in addition, reflection-positivity-like arguments allow for further limitations on the component functions as some of them are strictly positive, or strictly negative, or form a symmetric matrix of correlation functions with positive eigenvalues.
Also, the implications of EMC restrict the component functions even further.
These constraints can be tested on the data to establish limitations from lattice artifacts, gradient flow, or statistical uncertainties.
And as shown in~\cref{sec:SD}, the set of component functions can be described by an even smaller set of spectral functions.
In a lattice study, the correlation functions exhibit a bad signal-to-noise ratio for large separations.
As mentioned in the introduction, the authors from~\cite{Altenkort:2021jbk,Altenkort:2022yhb} tackle this issue by performing a functional fit in the large-$r$ region of the correlator.
Knowing that all component functions are described by a smaller set of spectral functions allows for a simultaneous fit of all component functions with a smaller set of fit parameters, representing the lightest state in each spectral function.

However, we have not yet explained how to perform step \ref{itm:app_last_step} from the list above, so it will be the main subject of this section.
In the next \namecrefs{subsec:app_transport_coefficients}, we explore how these component functions $G^{X}$ actually show up in the calculation of the relevant correlators.
We also briefly examine what happens at vanishing separation.

\subsection{Calculation of Transport Coefficients}\label{subsec:app_transport_coefficients}

A first transport coefficient to explore is the shear viscosity $\eta$, which is related via a Kubo formula to a real-time EMT correlation function \cite{Hosoya:1983id}, which itself can be obtained by analytic continuation from the space-averaged Euclidean time EMT correlation function
\cite{Karsch:1986cq}
\begin{align}\label{eq:app_A_eta_definition}
	A^{\eta}(\tau)&\coloneqq\frac{1}{10}\int\d*[3]{r}\ev{\pi_{ij}(\tau,\vec{r})\pi_{ij}(0,\vec{0})} &
	\pi_{ij}&\coloneqq T_{ij}-\frac{1}{3}\delta_{ij}T_{kk}
\end{align}
where the sum over repeated spatial indices $i,j,k$ is implied.
This correlator of purely spatial, traceless EMT components can be written via a single projector $P^{\eta}$ as
\begin{align}
	A^{\eta}(\tau)&=\frac{1}{10}\int\d*[3]{r}P^{\eta}_{\mu\nu\rho\sigma}(\hat{\vec{r}})G_{\mu\nu\rho\sigma}(\tau,\vec{r})
\end{align}
with
\begin{align}
	P^{\eta}_{\mu\nu\rho\sigma}(\hat{\vec{r}})
	&\coloneqq\frac{1}{2}\left(
		\delta^{\ispace}_{\mu\rho}\delta^{\ispace}_{\nu\sigma}
		+\delta^{\ispace}_{\mu\sigma}\delta^{\ispace}_{\nu\rho}
		-\frac{2}{3}\delta^{\ispace}_{\mu\nu}\delta^{\ispace}_{\rho\sigma}
	\right)
	=P^{\Trans\Trans}_{\mu\nu\rho\sigma}+P^{\Longi\Trans}_{\mu\nu\rho\sigma}+P^{\longi\longi}_{\mu\nu\rho\sigma},
\end{align}
where the projector has been expressed in terms of the projectors from~\cref{subsec:TD_fT_gTau}.
The action of the latter projectors on the full correlation function evaluates to the component functions times the multiplicity of the projector and hence one gets
\begin{align}
	\label{eq:app_A_eta_result}
	A^{\eta}(\tau)&=4\pi\int\d*{r}r^{2}\left[
		\frac{1}{5}G^{\Trans\Trans}(\tau,r^{2})
		+\frac{1}{5}G^{\Longi\Trans}(\tau,r^{2})
		+\frac{1}{10}G^{\longi\longi}(\tau,r^{2})
	\right].
\end{align}

This procedure can also be applied for the bulk viscosity $\zeta$, which is related to the Euclidean EMT correlation function
\begin{align}
	A^{\zeta}(\tau)&\coloneqq\int\d*[3]{r}\ev{T_{\mu\mu}(\tau,\vec{r})T_{\nu\nu}(0,\vec{0})},
\end{align}
which can again be expressed via a single projector $P^{\zeta}$ as
\begin{align}
	A^{\zeta}(\tau)&=\int\d*[3]{r}P^{\zeta}_{\mu\nu\rho\sigma}(\hat{\vec{r}})G_{\mu\nu\rho\sigma}(\tau,\vec{r}),
\end{align}
which itself is given by
\begin{align}
	P^{\zeta}&\coloneqq\delta_{\mu\nu}\delta_{\rho\sigma}
	=4P^{\scal\scal}_{\mu\nu\rho\sigma},
\end{align}
and hence the correlation function is given by
\begin{align}
	A^{\zeta}(\tau)=16\pi\int\d*{r}r^{2}G^{\scal\scal}(\tau,r^{2}).
\end{align}

Kovtun and Shukla have shown \cite{Kovtun:2018dvd} that certain \enquote{thermodynamical} second-order transport coefficients can be computed in terms of the $\tau$-integrated, zero-momentum limits of momentum-differentiated EMT correlation functions.
For instance, in eq.~(2.22) of their paper, they define the quantity
\begin{align}
	f_{1} &\coloneqq -\frac{1}{2}\lim_{\vec{k}\to0} \frac{\partial^{2}}{\partial k_{z}^{2}} G_{xyxy}(\vec{k},k^{0}=0).
\end{align}
Fourier transforming to coordinate space, this can be re-expressed as
\begin{align}
	f_1 &= \frac{1}{2} \int_{0}^{\beta}\d*{\tau}\pullint\int\d*[3]{r} r_{z}^{2} G_{xyxy}(\tau,\vec{r}).
\end{align}
In fact, all of the functions which appear in their analysis can be similarly written as spacetime-integrated moments of $G_{\mu\nu\rho\sigma}$-components weighted by polynomials in the coordinate.
Defining
\begin{align}
	A_{\alpha\beta,\mu\nu\rho\sigma} &\coloneqq \int\d*{\tau}\pullint\int\d*[3]{r} r_{\alpha} r_{\beta} G_{\mu\nu\rho\sigma}(\tau,\vec{r}),
\end{align}
Kovtun and Shukla showed that all thermodynamical second-order transport coefficients determined by the EMT can be expressed in terms of four such quantities:
\begin{subequations}
	\begin{align}
		A_{zz,xyxy}
		,&&
		A_{zz,xtxt}
		,&&
		A_{zz,tttt}
		,&&
		A_{zz,xxtt}.
	\end{align}
\end{subequations}
These can be evaluated by expressing $G_{\mu\nu\rho\sigma}$ in terms of component functions and projectors using~\cref{eq:TD_fT_aTau_G} and then carrying out the angular integration.
As an example, the first relation expands into
\begin{subequations}
	\label{eq:app_z2_xyxy}
	\begin{align}
		A_{zz,xyxy}
		&=\int\d*{\tau}\pullint\int\d*[3]{r}r_{z}^{2}G_{xyxy}(\tau,\vec{r}) \\
		&=\int\d*[3]{r}r_{z}^{2}\left[
			P^{\Trans\Trans}_{xyxy}(\hat{\vec{r}}) G^{\Trans\Trans}(r^{2})
			+P^{\Longi\Trans}_{xyxy}(\hat{\vec{r}}) G^{\Longi\Trans}(r^{2})
			+P^{\longi\longi}_{xyxy}(\hat{\vec{r}}) G^{\longi\longi}(r^{2})
		\right] \\
		\begin{split}
			&=\int\d*[3]{r}r_{z}^{2}
			\begin{aligned}[t]
				\bigg[
					&\frac{1}{2}\left(1-\hat{r}_{x}^{2}-\hat{r}_{y}^{2}+\hat{r}_{x}^{2}\hat{r}_{y}^{2}\right)G^{\Trans\Trans}(r^{2}) \\
					&+\frac{1}{2}\left(\hat{r}_{x}^{2}+\hat{r}_{y}^{2}-4\hat{r}_{x}^{2}\hat{r}_{y}^{2}\right)G^{\Longi\Trans}(r^{2})
					+\frac{3}{2}\hat{r}_{x}^{2}\hat{r}_{y}^{2}G^{\longi\longi}(r^{2})
				\bigg]
			\end{aligned}
		\end{split} \\
		&=4\pi\int\d*{r}r^{4}\left[
			\frac{11}{105}G^{\Trans\Trans}(r^{2})
			+\frac{1}{21}G^{\Longi\Trans}(r^{2})
			+\frac{1}{70}G^{\longi\longi}(r^{2})
		\right].
	\end{align}
\end{subequations}
The other three quantities can be derived in the same way and are given by
\begin{align}
	\label{eq:app_z2_xtxt}
	A_{zz,xtxt}&=4\pi\int\d*{r}r^{4}\left[
		\frac{2}{15}G^{\iTime\Trans}(r^{2})
		+\frac{1}{30}G^{\iTime\Longi}(r^{2})
	\right]
	\\
	\label{eq:app_z2_tttt}
	A_{zz,tttt}&=4\pi\int\d*{r}r^{4}\left[
		\frac{1}{12}G^{\scal\scal}(r^{2})
		+\frac{1}{4}G^{\enth\enth}(r^{2})
		+\frac{1}{2\sqrt{3}}G^{\scal\enth}(r^{2})
	\right]
	\\
	\label{eq:app_z2_xxtt}
	\begin{split}
		A_{zz,xxtt}&=4\pi\int\d*{r}r^{4}
		\begin{aligned}[t]
			\bigg[
				\frac{1}{12}G^{\scal\scal}(r^{2})
				&-\frac{1}{12}G^{\enth\enth}(r^{2})
				-\frac{1}{15\sqrt{6}}G^{\scal\longi}(r^{2})
				\\
				&+\frac{1}{6\sqrt{3}}G^{\scal\enth}(r^{2})
				-\frac{1}{15\sqrt{2}}G^{\longi\enth}(r^{2})
			\bigg].
		\end{aligned}
	\end{split}
\end{align}
In~\cref{eq:app_z2_xyxy,eq:app_z2_xtxt,eq:app_z2_tttt,eq:app_z2_xxtt}, the time-integrated component functions $G^{X}(r^{2})$ are used.

Note that, for shear viscosity, one often sees  $A^\eta(\tau) = \int\d[3]{r} \: G_{xyxy}(\tau,\vec r)$ instead of \eqref{eq:app_A_eta_definition}.
This amounts to computing one component of the component-average in that equation.
Applying angular averaging as in \eqref{eq:app_z2_xyxy} indeed returns \eqref{eq:app_A_eta_result}.

\subsection{Vanishing Spatial Separation}\label{subsec:app_vanishing_spatial_separation}
Because the EMT is dimension D, one expects the small-separation limit of the coefficient functions to scale as $G^{X}(r^{2})\propto r^{-2D}$.
Furthermore, because derivatives of $\hat{\vec{r}}$ scale as $1/r$, the projectors become ever faster varying.
Therefore, the $\vec{r}\to 0$ limits, both of the component functions and of the projectors, are singular.
However, most modern lattice calculations of EMT correlators utilize \textsl{gradient flow}
\cite{Luscher:2009eq,Luscher:2010iy}
to renormalize operators and eliminate such singular behavior.
This renders the correlation function finite and regular as $r\to 0$.
(Note that it also destroys the differential relations implied by EMC, since contact terms at $r=0$ are smeared into a region around $r=0$.)
On the lattice, the zero-separation point is a lattice site and represents a finite (though small) fraction of the lattice volume, so we should have some prescription for defining the correlator at zero separation.
And when considering $\tau$-dependent correlation functions, the finite-$\tau$ but zero-$\vec r$ points must certainly be considered.
Therefore we should find some definition for the projectors at vanishing spatial separation, and it makes sense to define them at vanishing separation in terms of their angular average,
i.e.\
\begin{align}
	P^{X}_{\mu\nu\rho\sigma}(\hat{\vec{r}})&\xrightarrow{r\to0}\int\frac{\d\Omega}{A_D}P^{X}_{\mu\nu\rho\sigma}(\hat{\vec{r}}),
	&
	A_{D}&=
	\begin{cases}
		4\pi & D=3 \\
		2\pi^{2} & D=4
	\end{cases}.
\end{align}

For the zero temperature case, the projectors $P^{X}$ for $X\in\set{\longi\longi,\Longi\Trans,\Trans\Trans}$ become
\begin{align}\label{eq:app_VS_r_to_0}
	P^{X}_{\mu\nu\rho\sigma} &\xrightarrow{r\to 0} \frac{M^{X}}{2(M^{\longi\longi}+M^{\Longi\Trans}+M^{\Trans\Trans})} \left(
		\delta_{\mu\rho}\delta_{\nu\sigma}
		+\delta_{\mu\sigma}\delta_{\nu\rho}
		-\frac{2}{D}\delta_{\mu\nu}\delta_{\rho\sigma}
	\right)
\end{align}
while $P^{\scal\scal}$ is unchanged and $P^{\scal\longi}$ vanishes.
As expected, the zero-separation limits of the orthogonal projectors $P^{\Trans\Trans}$, $P^{\Longi\Trans}$, $P^{\scal\scal}$ and $P^{\longi\longi}$ add up to the identity projector $P^{\all}$, see~\cref{eq:TD_zT_Pall}.
Because several projectors become degenerate in this limit, this implies that $G^{\Trans\Trans}(\vierervec{0})=G^{\Longi\Trans}(\vierervec{0})=G^{\longi\longi}(\vierervec{0})$ and $G^{\scal\longi}(\vierervec{0})=0$.

In the scenario of finite temperature and a finite temporal separation, the fourteen projectors have the following zero-spatial-separation behavior:
\begin{subequations}\label{eq:app_VS_fT_gTau_projectors}
	\begin{align}
		P^{\Trans\Trans}_{\mu\nu\rho\sigma}\xrightarrow{r\to0}{}
		&\frac{1}{5}\left(
			\delta^{\ispace}_{\mu\rho}\delta^{\ispace}_{\nu\sigma}
			+\delta^{\ispace}_{\mu\sigma}\delta^{\ispace}_{\nu\rho}
			-\frac{2}{3}\delta^{\ispace}_{\mu\nu}\delta^{\ispace}_{\rho\sigma}
		\right)
		\\
		P^{\Longi\Trans}_{\mu\nu\rho\sigma}\xrightarrow{r\to0}{}
		&\frac{1}{5}\left(
			\delta^{\ispace}_{\mu\rho}\delta^{\ispace}_{\nu\sigma}
			+\delta^{\ispace}_{\nu\rho}\delta^{\ispace}_{\mu\sigma}
			-\frac{2}{3}\delta^{\ispace}_{\mu\nu}\delta^{\ispace}_{\rho\sigma}
		\right)
		\\
		P^{\longi\longi}_{\mu\nu\rho\sigma}\xrightarrow{r\to0}{}
		&\frac{1}{10}\left(
			\delta^{\ispace}_{\mu\rho}\delta^{\ispace}_{\nu\sigma}
			+\delta^{\ispace}_{\mu\sigma}\delta^{\ispace}_{\nu\rho}
			-\frac{2}{3}\delta^{\ispace}_{\mu\nu}\delta^{\ispace}_{\rho\sigma}
		\right)
		\\
		P^{\iTime\Trans}_{\mu\nu\rho\sigma}\xrightarrow{r\to0}{}
		&\frac{1}{3}\left(
			u_{\mu}u_{\rho}\delta^{\ispace}_{\nu\sigma}
			+u_{\mu}u_{\sigma}\delta^{\ispace}_{\nu\rho}
			+u_{\nu}u_{\rho}\delta^{\ispace}_{\mu\sigma}
			+u_{\nu}u_{\sigma}\delta^{\ispace}_{\mu\rho}
		\right)
		\\
		P^{\mix\mix}_{\mu\nu\rho\sigma}\xrightarrow{r\to0}{}
		&\frac{1}{6}\left(
			u_{\mu}u_{\rho}\delta^{\ispace}_{\nu\sigma}
			+u_{\mu}u_{\sigma}\delta^{\ispace}_{\nu\rho}
			+u_{\nu}u_{\rho}\delta^{\ispace}_{\mu\sigma}
			+u_{\nu}u_{\sigma}\delta^{\ispace}_{\mu\rho}
		\right)
		\\
		P^{\enth\enth}_{\mu\nu\rho\sigma}\xrightarrow{r\to0}{}
		&\frac{1}{12}\left(
			9u_{\mu}u_{\nu}u_{\rho}u_{\sigma}
			-3u_{\rho}u_{\sigma}\delta^{\ispace}_{\mu\nu}
			-3u_{\mu}u_{\nu}\delta^{\ispace}_{\rho\sigma}
			+\delta^{\ispace}_{\mu\nu}\delta^{\ispace}_{\rho\sigma}
		\right)
		\\
		P^{\scal\enth}_{\mu\nu\rho\sigma}\xrightarrow{r\to0}{}
		&\frac{1}{\sqrt{12}}\left(
			3u_{\mu}u_{\nu}u_{\rho}u_{\sigma}
			+u_{\mu}u_{\nu}\delta^{\ispace}_{\rho\sigma}
			+u_{\rho}u_{\sigma}\delta^{\ispace}_{\mu\nu}
			-\delta^{\ispace}_{\mu\nu}\delta^{\ispace}_{\rho\sigma}
		\right)
		\\
		P^{\scal\scal}_{\mu\nu\rho\sigma}\xrightarrow{r\to0}{}
		&\frac{1}{4}\left(
			u_{\mu}u_{\nu}u_{\rho}u_{\sigma}
			+u_{\mu}u_{\nu}\delta^{\ispace}_{\rho\sigma}
			+u_{\rho}u_{\sigma}\delta^{\ispace}_{\mu\nu}
			+\delta^{\ispace}_{\mu\nu}\delta^{\ispace}_{\rho\sigma}
		\right)
	\end{align}
	while the others vanish
	\begin{align}
		P^{\scal\longi}_{\mu\nu\rho\sigma},
		P^{\Mix\Trans}_{\mu\nu\rho\sigma},
		P^{\scal\mix}_{\mu\nu\rho\sigma},
		P^{\longi\enth}_{\mu\nu\rho\sigma},
		P^{\longi\mix}_{\mu\nu\rho\sigma},
		P^{\enth\mix}_{\mu\nu\rho\sigma}\xrightarrow{r\to0}{}&0.
	\end{align}
\end{subequations}
Therefore $G^{\Trans\Trans}(\tau,\vec{0})=G^{\Longi\Trans}(\tau,\vec{0})=G^{\longi\longi}(\tau,\vec{0})$ and
$G^{\iTime\Trans}(\tau,\vec{0})=G^{\mix\mix}(\tau,\vec{0})$, due to the respective multiplicities.
In general this means that there are five distinct nonzero elements at zero spatial separation.

\section{Discussion} \label{sec:Discussion}

In this purely technical paper, we have explored how the Euclidean EMT two-point correlation function can be decomposed into a minimal basis of tensorial structures times coefficient functions for each structure, and we have explored the inter-relation between those coefficient functions implied by energy-momentum conservation.
We also showed how to use these results in lattice studies, giving a prescription for how to decompose lattice-determined EMT correlators into the component functions we have defined.
Most significantly, for applications to transport, we showed how each relevant transport coefficient is determined in terms of radial integrals of specific component functions.
This will allow lattice practitioners to use our decomposition technique for the evaluation off all of the transport coefficients discussed in the previous section (and potentially others which have not yet been identified).
Work towards carrying out such lattice studies is already ongoing.

Of course, such investigations have a long history, so we should emphasize how our approach improves on the status quo.
To do so we will focus on a specific example, the determination of $A_{zz,xyxy}$ for the evaluation of the thermodynamical transport coefficient $\kappa$.
Philipsen and Sch\"afer made a pioneering investigation of this correlator in pure-glue QCD in Ref.~\cite{Philipsen:2013nea}.
Without the tools we have developed, they did so by directly evaluating the position-space correlator
$\int \d{x}\d{y}\d{\tau} \langle T_{xy}(x,y,z,\tau) T_{xy}(0,0,0,0) \rangle$ as a function of $z$ by integrating each stress-tensor operator over the lattice $(x,y,\tau)$ plane.
They then investigated the $z$-dependence of the resulting correlator and used Fourier methods to perform $\lim_{k_z \to 0} k_z^2 G_{xyxy}(k_z)$.
The problem with this approach is that one is forced to integrate over all $(x,y)$ separations.
Either one must choose a lattice where the $(x,y)$ extent is small, introducing finite-volume effects, or one must include large transverse separations.
But, as shown by Parisi \cite{Parisi:1983ae}, large-separation correlators exhibit fluctuations as severe as short-distance correlators, so including them includes noise without any signal.
Therefore such an approach necessarily either introduces finite-volume effects or suffers from signal-to-noise problems.
In pure-glue QCD these can be alleviated by the use of multilevel methods, but these are not available in full QCD.
In contrast, under our approach one uses all components of the stress tensor in all separation directions to determine $G^{\Trans\Trans},G^{\Longi\Longi},G^{\longi\longi}$ as a function of radius.
This uses all data and it allows us to cut off the separations considered at large radius, where the data becomes noisy but the functional form of the correlator is known based on the arguments we give above.
This allows the large-$r$ tail to be fit and the noisy large-$r$ data to be replaced by this fit.
The use of all available lattice correlators and the elimination of noisy large-separation data significantly improves the signal-to-noise ratio.
Similar but somewhat less elegant methods, applied to shear viscosity, have already been used to improve signal-to-noise by about a factor of 7
\cite{Altenkort:2021jbk,Altenkort:2022yhb,Ding:2026gco}.
Therefore, we believe that the prospects for high-precision evaluations of the thermodynamical transport coefficients, both in pure-glue and in full QCD, are excellent if the tools developed here are used.
We also see our approach as the only realistic way to get good quality lattice data in full QCD for the reconstruction of the shear and bulk viscosities.

\section*{Acknowledgments}

We would like to thank Sangyong Jeon for useful conversations and Larry Yaffe for sharing (indirectly) unpublished notes about the decomposition in momentum space.
We also thank Pavan, Rasmus Larsen, and Olaf Kaczmarek for discussions and collaboration on the forthcoming applications to lattice gauge theory.
This research was funded by the DFG (Collaborative Research Center CRC-TR 211 ``Strong-interaction matter under
extreme conditions''~--~project number \mbox{315477589~--~TRR 211}).

\newpage
\bibliography{references.bib}

\end{document}